\documentclass[sigconf,natbib=false,nonacm]{acmart}

\usepackage[utf8]{inputenc}
\usepackage[T1]{fontenc}
\usepackage{stmaryrd}
\usepackage{algorithmic}
\usepackage{graphicx}
\usepackage{tabularx}
\usepackage{booktabs}
\usepackage{textcomp}
\usepackage{xcolor}
\usepackage[style=acmnumeric,sorting=none]{biblatex}
\usepackage[font=small]{subfig}

\usepackage[scaled=.8]{beramono}
\usepackage{hyperref}
\usepackage[capitalize,nameinlink]{cleveref}
\usepackage{xspace}
\usepackage[inline]{enumitem}
\newlist{andlist}{enumerate*}{1}
\setlist[andlist]{itemjoin={{, }},itemjoin*={{, and }},label=(\(\roman*\)),mode=boxed}
\usepackage{tikz}
\usepackage{microtype}
\usepackage{fontawesome5}
\usetikzlibrary{shapes.misc,arrows.meta,tikzmark,decorations.pathmorphing,decorations.pathreplacing,positioning,matrix}
\allowdisplaybreaks

\usepackage{xfp}
\usepackage[detect-all]{siunitx}
\DeclareSymbolFont{AMSbb}{U}{pxsyb}{m}{n}
\SetSymbolFont{AMSbb}{bold}{U}{pxsyb}{bx}{n}
\DeclareFontSubstitution{U}{pxsyb}{m}{n}
\DeclareSymbolFontAlphabet{\mathbb}{AMSbb}
\DeclareMathAlphabet{\mathcal}{OMS}{cmsy}{m}{n}
\usepackage[print,numinpar,fakeminted]{lib/xlistings/xlistings}
\LoadLanguages{R}
\xlstsetstyle{plain number}
\xlstlangoverride{R}{
    deletekeywords={data,frame},
    otherkeywords={data.frame}
}
\lstcolorlet{keywordA}{black}
\lstcolorlet{keywordB}{black}
\lstcolorlet{keywordC}{black}
\lstcolorlet{literals}{black}
\lstcolorlet{numbers}{gray}

\xlstDefineStyles{%
    {linenumbers:  \color{gray}\scriptsize\sffamily}%
}
\crefname{lstlisting}{Listing}{Listings}
\Crefname{lstlisting}{Listing}{Listings}

\usepackage{tcolorbox}
\newtcolorbox[auto counter]{finding}{
    hyphenationfix,
    top=4pt,
    bottom=3pt,
    left*=0pt,
    right*=0pt,
    grow sidewards by=5pt,
    colback=gray!5,
    colframe=lightgray,
    boxrule=1pt,
    boxsep=0pt,
    arc=4pt,
    before upper={\textbf{Finding \thetcbcounter:}\space\ignorespaces}
}

\newcommand*{\LeftArrow}{\text{\BeginAccSupp{method=escape,ActualText={<-}}\(\leftarrow\)\EndAccSupp{}}}
\newcommand*{\RightArrow}{\text{\BeginAccSupp{method=escape,ActualText={->}}\(\rightarrow\)\EndAccSupp{}}}
\newcommand*{\DoubleLeftArrow}{\text{\BeginAccSupp{method=escape,ActualText={<<-}}\(\twoheadleftarrow\)\EndAccSupp{}}}
\newcommand*{\DoubleRightArrow}{\text{\BeginAccSupp{method=escape,ActualText={->>}}\(\twoheadrightarrow\)\EndAccSupp{}}}
\newcommand*{\ConceptEvalScripts}{4230}
\newcommand*{\ConceptEvalDfFunctions}{118}
\newcommand*{\ConceptEvalReadCsvFilePercent}{48}
\newcommand*{\ConceptEvalDataFrameFilePercent}{38}
\newcommand*{\ConceptEvalColNameAccessPercent}{78}
\newcommand*{\ConceptEvalColIndexAccessPercent}{22}
\newcommand*{\ConceptEvalRowNameAccessPercent}{0.02}
\newcommand*{\ConceptEvalRowIndexAccessPercent}{5.5}

\newcommand*{\TestsTotal}{1737}
\newcommand*{\TestsCoverage}{96}

\newcommand*{\CorrEvalScripts}{78}
\newcommand*{\CorrEvalCorrDf}{925}
\newcommand*{\CorrEvalCorrEmpty}{3077}
\newcommand*{\CorrEvalWrongDf}{1}
\newcommand*{\CorrEvalWrongEmpty}{358}
\newcommand*{\CorrEvalPoints}{\fpeval{\CorrEvalCorrDf + \CorrEvalCorrEmpty + \CorrEvalWrongDf + \CorrEvalWrongEmpty}}
\newcommand*{\CorrEvalDfPoints}{\fpeval{\CorrEvalCorrDf + \CorrEvalWrongEmpty}}
\newcommand*{\CorrEvalDfPointsPercent}{\fpeval{round(100 * \CorrEvalDfPoints / \CorrEvalPoints, 0)}}
\newcommand*{\CorrEvalDfCorrDfPercent}{\fpeval{round(100 * \CorrEvalCorrDf / \CorrEvalDfPoints, 0)}}
\newcommand*{\CorrEvalWrongEmptyPercent}{\fpeval{round(100 * \CorrEvalWrongEmpty / \CorrEvalDfPoints, 0)}}
\newcommand*{\CorrEvalOverApprox}{854}
\newcommand*{\CorrEvalTop}{488}
\newcommand*{\CorrEvalExact}{71}
\newcommand*{\CorrEvalOverApproxPercent}{\fpeval{round(100 * \CorrEvalOverApprox / \CorrEvalCorrDf, 0)}}
\newcommand*{\CorrEvalTopPercent}{\fpeval{round(100 * \CorrEvalTop / \CorrEvalCorrDf, 0)}}
\newcommand*{\CorrEvalExactPercent}{\fpeval{round(100 * \CorrEvalExact / \CorrEvalCorrDf, 0)}}

\newcommand*{\CorrEvalColNamesMidRelDiff}{-0.03717636147239083}
\newcommand*{\CorrEvalColCountMidRelDiff}{0.873493574154262}
\newcommand*{\CorrEvalRowCountMidRelDiff}{411.9013990910672}

\newcommand*{\CorrEvalColNamesMidDiffTimes}{\fpeval{round(1.0 + \CorrEvalColNamesMidRelDiff, 2)}}
\newcommand*{\CorrEvalColCountMidDiffTimes}{\fpeval{round(1.0 + \CorrEvalColCountMidRelDiff, 2)}}
\newcommand*{\CorrEvalRowCountMidDiffTimes}{\fpeval{round(1.0 + \CorrEvalRowCountMidRelDiff, 0)}}

\newcommand*{\TotalEvalScripts}{33314}
\newcommand*{\TotalEvalScriptsDatasets}{1437}
\newcommand*{\TotalEvalScriptsDatasetPercent}{\fpeval{round(100 * \TotalEvalScriptsDatasets / \TotalEvalScripts, 1)}}
\newcommand*{\TotalEvalFailed}{19}

\newcommand*{\TotalEvalSuccess}{\fpeval{\TotalEvalScripts - \TotalEvalFailed}}
\newcommand*{\TotalEvalDfScripts}{23964}
\newcommand*{\TotalEvalDfScriptsPercent}{\fpeval{round(100 * \TotalEvalDfScripts / \TotalEvalSuccess, 1)}}
\newcommand*{\TotalEvalAvgNumOperations}{70.9}
\newcommand*{\TotalEvalSdNumOperations}{227}

\newcommand*{\TotalEvalAvgConstraints}{21.4}
\newcommand*{\TotalEvalSdConstraints}{58.9}
\newcommand*{\TotalEvalTotalConstraints}{712505}
\newcommand*{\TotalEvalTotalFunctionCalls}{21101688}
\newcommand*{\TotalEvalDfCallsPercent}{\fpeval{round(100 * \TotalEvalTotalConstraints / \TotalEvalTotalFunctionCalls, 1)}}
\newcommand*{\TotalEvalDfShapeExact}{6659}
\newcommand*{\TotalEvalDfShapeValue}{4599}
\newcommand*{\TotalEvalDfShapePartial}{288575}
\newcommand*{\TotalEvalDfShapeTop}{412672}
\newcommand*{\TotalEvalDfShapeNotTop}{\fpeval{\TotalEvalTotalConstraints - \TotalEvalDfShapeTop}}
\newcommand*{\TotalEvalDfShapeExactPercent}{\fpeval{round(100 * \TotalEvalDfShapeExact / \TotalEvalTotalConstraints, 1)}}
\newcommand*{\TotalEvalDfShapeValuePercent}{\fpeval{round(100 * \TotalEvalDfShapeValue / \TotalEvalTotalConstraints, 1)}}
\newcommand*{\TotalEvalDfShapePartialPercent}{\fpeval{round(100 * \TotalEvalDfShapePartial / \TotalEvalTotalConstraints, 1)}}
\newcommand*{\TotalEvalDfShapeTopPercent}{\fpeval{round(100 * \TotalEvalDfShapeTop / \TotalEvalTotalConstraints, 1)}}
\newcommand*{\TotalEvalDfShapeNotTopPercent}{\fpeval{round(100 * \TotalEvalDfShapeNotTop / \TotalEvalTotalConstraints, 1)}}
\newcommand*{\TotalEvalDatasetsConstraints}{41264}
\newcommand*{\TotalEvalDatasetsExact}{1713}
\newcommand*{\TotalEvalDatasetsTop}{17057}
\newcommand*{\TotalEvalDatasetsNotTop}{\fpeval{\TotalEvalDatasetsConstraints - \TotalEvalDatasetsTop}}
\newcommand*{\TotalEvalDatasetsExactPercent}{\fpeval{round(100 * \TotalEvalDatasetsExact / \TotalEvalDatasetsConstraints, 1)}}
\newcommand*{\TotalEvalDatasetsNotTopPercent}{\fpeval{round(100 * \TotalEvalDatasetsNotTop / \TotalEvalDatasetsConstraints, 1)}}
\newcommand*{\TotalEvalColNamesRefine}{2483}
\newcommand*{\TotalEvalColCountRefine}{907218}
\newcommand*{\TotalEvalRefine}{\fpeval{\TotalEvalColNamesRefine + \TotalEvalColCountRefine}}
\newcommand*{\TotalEvalNoRefine}{370097}
\newcommand*{\TotalEvalRefinePercent}{\fpeval{round(100 * \TotalEvalRefine / (\TotalEvalRefine + \TotalEvalNoRefine), 0)}}
\newcommand*{\TotalEvalWidenings}{3829}
\newcommand*{\TotalEvalVectorArgValue}{127040}
\newcommand*{\TotalEvalVectorArgLength}{47847}
\newcommand*{\TotalEvalVectorArgTop}{219828}
\newcommand*{\TotalEvalStringArgValue}{55237}
\newcommand*{\TotalEvalStringArgTop}{11927}
\newcommand*{\TotalEvalNumberArgValue}{70166}
\newcommand*{\TotalEvalNumberArgTop}{174267}
\newcommand*{\TotalEvalVectorArgTopPercent}{\fpeval{round(100 * \TotalEvalVectorArgTop / (\TotalEvalVectorArgValue + \TotalEvalVectorArgLength + \TotalEvalVectorArgTop), 0)}}
\newcommand*{\TotalEvalStringArgTopPercent}{\fpeval{round(100 * \TotalEvalStringArgTop / (\TotalEvalStringArgValue + \TotalEvalStringArgTop), 0)}}
\newcommand*{\TotalEvalNumberArgTopPercent}{\fpeval{round(100 * \TotalEvalNumberArgTop / (\TotalEvalNumberArgValue + \TotalEvalNumberArgTop), 0)}}
\newcommand*{\TotalEvalAvgInferenceTime}{\qty{163}{\milli\second}\xspace}
\newcommand*{\TotalEvalSdInferenceTime}{\qty{1.8}{\second}\xspace}
\newcommand*{\TotalEvalMedianInferenceTime}{\qty{49}{\milli\second}\xspace}
\newcommand*{\TotalEvalAvgTotalTime}{\qty{826}{\milli\second}\xspace}
\newcommand*{\TotalEvalSdTotalTime}{\qty{2.6}{\second}\xspace}
\newcommand*{\TotalEvalMedianTotalTime}{\qty{456}{\milli\second}\xspace}

\newcommand*{\LinterEvalFailed}{33}

\newcommand*{\LinterEvalSuccess}{\fpeval{\TotalEvalScripts - \LinterEvalFailed}}
\newcommand*{\LinterEvalIssues}{2158}
\newcommand*{\LinterEvalIssueScripts}{275}
\newcommand*{\LinterEvalIssueScriptsPercent}{\fpeval{round(100 * \LinterEvalIssueScripts / \LinterEvalSuccess, 2)}}
\newcommand*{\LinterEvalColNamesIssues}{1729}
\newcommand*{\LinterEvalColIndexIssues}{429}

\newcommand*{\LinterEvalColNamesIssuesPercent}{\fpeval{round(100 * \LinterEvalColNamesIssues / \LinterEvalIssues, 0)}}
\newcommand*{\LinterEvalColIndexIssuesPercent}{\fpeval{round(100 * \LinterEvalColIndexIssues / \LinterEvalIssues, 0)}}

\newcommand*{\LinterEvalActualIssueScripts}{40}
\newcommand*{\LinterEvalFalsePositiveScripts}{\fpeval{\LinterEvalIssueScripts - \LinterEvalActualIssueScripts}}

\newcommand*{\LinterEvalActualWrongDatasetScripts}{11}
\newcommand*{\LinterEvalActualWrongCapitalizationScripts}{4}

\setcopyright{none}
\acmConference[]{}{}{}
\acmBooktitle{}
\acmDOI{}
\acmISBN{}
\copyrightyear{}
\renewcommand\footnotetextcopyrightpermission[1]{}
\begin{abstract}
Data frames are a fundamental data structure in many data analysis tasks and are widely used in programming languages like~R. Due to their omnipresence in data analysis, there are many functions that operate on their shape and content, for example, to clean and transform study data. However, languages like~R do not offer static guarantees on data frames making it difficult to reason about their shape at a specific point in the program.

In this paper, we present a novel static analysis to infer the shape of data frames in R~programs using abstract interpretation by tracking the ensured and potential column names, as well as the potential number of columns and rows. For this, we use a reduced product domain and define abstract semantics for the most commonly used data frame operations, such as mutating, filtering, and subsetting. We evaluate the correctness and accuracy of our analysis on a selection of \num{\CorrEvalScripts}~executable real-world R~scripts achieving empirical evidence for soundness by never under-approximating the data frame shape. Additionally, we demonstrate the ability of our analysis to infer the shape of data frames on a large dataset of \num{\TotalEvalScripts}~real-world R~scripts by inferring \emph{concrete} shape constraints for \qty{\TotalEvalDfShapeNotTopPercent}{\percent}~and \emph{exact} shapes for~\qty{\TotalEvalDfShapeExactPercent}{\percent}~of the data frame operations, improving to \qty{\TotalEvalDatasetsNotTopPercent}{\percent} and \qty{\TotalEvalDatasetsExactPercent}{\percent} if all datasets read in these scripts are available to our analysis. Using the inferred data frame shapes, we identified \num{\LinterEvalActualIssueScripts}~real-world R~scripts containing potential invalid data frame accesses. This shows the potential of our analysis to significantly support researchers in using data frames in data analysis.
\end{abstract}

\ccsdesc[500]{Software and its engineering~Automated static analysis}
\ccsdesc[500]{Theory of computation~Abstraction}

\keywords{Abstract Interpretation, Static Analysis, R~Language, Data Frames}

\begin{document}

\newcommand*{\code}[1]{{\normalfont\texttt{\frenchspacing#1}}}
\newcommand*{\codedots}{\code{.\!.\!.}}
\newcommand*{\assign}{\LeftArrow\xspace}
\newcommand*{\flowR}{\emph{flowR}\xspace} 

\newcounter{markercounter}
\newcommand*{\marker}[1]{\tikzmarknode{#1}{\strut}}
\newcommand*{\marktext}[2]{\tikzmarknode{#1}{#2}}
\newcommand*{\textbrace}[3][]{%
	\def\markername{marker\arabic{markercounter}}\stepcounter{markercounter}%
	\marker{\markername-1}{#2}\marker{\markername-2}%
    \tikz[overlay,remember picture,line cap=round,line join=round,gray,font=\tiny]%
    \draw[decorate,decoration={brace,mirror}] (\markername-1.south) to[edge node={node[below=1pt,align=center,#1] {\vphantom{l}#3}}] (\markername-2.south);%
}

\newcommand*{\xlongleftrightarrows}[2]{\mathrel{\raise1.5ex\hbox{\oalign{%
  \(\scriptstyle\hfil\smash#1\hfil\)\cr%
  \(\scriptstyle\leftarrow\joinrel\relbar\joinrel\relbar\hfil\)\cr%
  \vrule width0pt height.25ex%
  \(\scriptstyle\hfil\relbar\joinrel\relbar\joinrel\smash\rightarrow\)\cr%
  \(\scriptstyle\hfil#2\hfil\)\cr%
}}}}

\newcommand*{\abstractsym}{\#}
\newcommand*{\absintsub}[1]{{\kern-0.25pt\scalebox{0.6}{\ensuremath{\mathcal{\uppercase{#1}}}}}}
\newcommand*{\absintsym}[2][]{\smash{{#2}^{\abstractsym}_\absintsub{#1}}}
\newcommand*{\absintcmd}[1]{\ensuremath{#1}\xspace}
\newcommand*{\df}{\absintcmd{d\!f}}
\newcommand*{\X}[1][]{\absintcmd{\absintsym[#1]{X}}}
\newcommand*{\DF}{\absintcmd{\absintsym{\mathcal{D\!F}}}}
\renewcommand*{\N}{\absintcmd{\absintsym{\mathcal{N}}}}
\renewcommand*{\C}{\absintcmd{\absintsym{\mathcal{C}}}}
\newcommand*{\R}{\absintcmd{\absintsym{\mathcal{R}}}}
\let\mathbot\bot \renewcommand*{\bot}{\absintcmd{\mathbot}}
\let\mathtop\top \renewcommand*{\top}{\absintcmd{\mathtop}}
\newcommand*{\topN}{\absintcmd{\top_{\kern-1pt\sigma}}}
\newcommand*{\rhoDF}{\absintcmd{\rho_\absintsub{d\!f}}}
\newcommand*{\lesseq}{\absintcmd{\sqsubseteq}}
\newcommand*{\join}{\absintcmd{\sqcup}}
\newcommand*{\meet}{\absintcmd{\sqcap}}
\newcommand*{\widen}{\absintcmd{\mathop{\triangledown}}}
\newcommand*{\narrow}{\absintcmd{\mathop{\triangle}}}
\newcommand*{\concretization}[1][]{\absintcmd{\gamma_\absintsub{#1}}}
\newcommand*{\abstraction}[1][]{\absintcmd{\alpha_\absintsub{#1}}}
\newcommand*{\galois}[2]{\absintcmd{\smash{\xlongleftrightarrows{#1}{#2}}}}
\newcommand*{\stat}[3]{\absintcmd{\absintsym[#1]{\mathbb{S}}\llbracket#2\rrbracket#3}}
\newcommand*{\expr}[3]{\absintcmd{\absintsym[#1]{\mathbb{E}}\llbracket#2\rrbracket#3}}
\newcommand*{\cond}[3]{\absintcmd{\absintsym[#1]{\mathbb{C}}\llbracket#2\rrbracket#3}}
\newcommand*{\powerset}[1]{\absintcmd{\mathcal{P}(#1)}}
\newcommand*{\strings}{\absintcmd{\Sigma^*}}
\newcommand*{\naturals}{\absintcmd{\mathbb{N}_0}}
\newcommand*{\setrange}[1]{\absintcmd{({#1})}}
\newcommand*{\interval}[1]{\absintcmd{[{#1}]}}
\newcommand*{\dotcup}{\absintcmd{\mathop{\dot{\cup}}}}
\newcommand*{\dotcap}{\absintcmd{\mathop{\dot{\cap}}}}

\newcommand*{\lref}[1]{\hyperref[#1]{line~\ref*{#1}}}
\newcommand*{\todo}[1]{\textcolor{purple}{\sffamily\small\faCaretRight~\textbf{TODO:}~#1~\faCaretLeft}}

\title{Inferring the Shape of Data~Frames in~R~Programs using~Abstract~Interpretation}

\author{Oliver Gerstl}
\email{oliver.gerstl@uni-ulm.de}
\orcid{0009-0007-5612-0780}
\affiliation{%
  \institution{Ulm University}
  \country{Germany}
}

\author{Florian Sihler}
\email{florian.sihler@uni-ulm.de}
\orcid{0000-0001-7195-7801}
\affiliation{%
  \institution{Ulm University}
  \country{Germany}
}

\author{Matthias Tichy}
\email{matthias.tichy@uni-ulm.de}
\orcid{0000-0002-9067-3748}
\affiliation{%
  \institution{Ulm University}
  \country{Germany}
}

\maketitle

\section{Introduction}\label{sec:introduction}

Programming languages like R, Python, and Julia are widely used in statistical computing and data analysis~\cite{ref:10.5555/3086927}. Hence, they are mostly used by researchers and data scientists that do not have a background in software engineering~\cite{ref:conf/wcre/0001WD18,ref:conf/vl/0002HB20}. Several studies show that research codes suffers from reproducibility issues because of bad coding practices~\cite{ref:journals/corr/abs-2103-12793,ref:conf/msr/IslamAW24}. Moreover, these languages incorporate powerful reflective features, such as the evaluation of code from strings and the modification of function bodies at runtime~\cite{ref:conf/ecoop/MorandatHOV12,ref:conf/msr/SihlerPSTDD24}.

One of the most commonly used data structures in data science programs is the data frame, a tabular data structure for storing, transforming, and visualizing data~\cite{ref:10.5555/3086927}. However, dynamic programming languages like~R do not provide a way to statically verify the shape of data frames, for example, to guarantee the existence of accessed columns and rows of data frames. This may lead to subtle errors due to missing column names or out-of-bounds accesses for columns and rows of data frames when running the code. This is particularly important for programs performing long-running analyses where a dynamic analysis is not feasible.

There are many static analysis tools for popular programming languages like Java and C/C++~\cite{ref:fbInfer24,ref:clanganalyzer25,ref:framaC25}. However, there is a lack of sufficient static analysis tools to support researchers writing data science programs~\cite{ref:Vidoni021,ref:10.1145/3763087} and using data science-specific data structures like data frames. Existing work on analyzing data frames focuses solely on Python and does not evaluate the effectiveness of their approach on real-world code~\cite{ref:conf/pldi/NegriniSU23,ref:journals/access/ZhuangL22}.

In this paper, we propose an approach to statically infer the shape of data frames in R~programs using abstract interpretation~\cite{ref:conf/popl/CousotC77} to aid researchers in writing data science code. Our contributions are a formal definition of our approach and an accompanying implementation. Additionally, to the best of our knowledge, we are the first to analyze data frames that evaluate this approach on real-world code. We define the following research questions to guide this paper:

\begin{enumerate}[label={RQ\arabic*:},ref={RQ\arabic*},leftmargin=*]
    \item\label{rq1} How can the shape of data frames be inferred using abstract interpretation?
    \item\label{rq2} How effectively can data frame shapes be inferred in real-world R~scripts?
\end{enumerate}

To answer~\ref{rq1}, we define an abstract domain for the shape of data frames as a partially reduced product domain for the column names, number of columns, and number of rows. We then describe the abstract semantics of commonly used data frame operations by mapping them to abstract data frame operations and defining their abstract semantics with respect to the data frame shape domain. We provide a companion implementation of our approach based on \flowR~\cite{ref:10.1145/3763087}, a static data flow analyzer for~R, and implement a linter rule to report invalid data frame accesses in R~programs.

Addressing \ref{rq2}, we perform a correctness evaluation of the data frame shape inference based on ground truth from \num{\CorrEvalScripts}~executable real-world R~scripts~\cite{ref:10.5281/zenodo.18927873}. We successfully identified \qty{\CorrEvalDfCorrDfPercent}{\percent}~of the actual data frames as such and did not under-approximate the actual data frame shape in any case, providing empirical evidence for soundness. Moreover, evaluating our approach on \num{\TotalEvalScripts}~real-world R~scripts~\cite{ref:10.5281/zenodo.18927873}, we inferred \emph{concrete} data frame shapes for~\qty{\TotalEvalDfShapeNotTopPercent}{\percent} and \emph{exact} data frame shapes for \qty{\TotalEvalDfShapeExactPercent}{\percent}~of the identified data frame operations, which improves to \qty{\TotalEvalDatasetsNotTopPercent}{\percent} and \qty{\TotalEvalDatasetsExactPercent}{\percent}, respectively, if all datasets read in the scripts are available to our analysis. Using the inferred data frame shapes, we found \num{\LinterEvalActualIssueScripts}~real-world R~scripts containing potential invalid data frame accesses. On average, the data frame shape inference took~\TotalEvalAvgInferenceTime per script with an average total runtime of~\TotalEvalAvgTotalTime including all preceding analysis steps. Although we focus on data frames in~R in this paper, the general approach of the reduced product domain for the shape of data frames can be generalized to other programming languages that provide tabular data structures like Python.

In \cref{sec:related_work}, we first discuss related work on static analysis approaches for data science and especially data frames. Second, we introduce a motivating example in \cref{sec:motivation} that we use as running example throughout this paper. In \cref{sec:approach}, we present our approach of statically inferring the shape of data frames and discuss our accompanying implementation. Subsequently, we evaluate the correctness and effectiveness of our approach in \cref{sec:evaluation} and conclude our work in \cref{sec:conclusion}.

\section{Related Work}\label{sec:related_work}

Abstract interpretation is a static analysis technique that soundly over-approximates the behavior of a program by abstracting program semantics~\cite{ref:conf/popl/CousotC77}. It is used in program verification~\cite{ref:conf/nfm/Cousot12,ref:conf/esop/CousotCFMMMR05}, contract checking~\cite{ref:conf/foveoos/FahndrichL10}, type inference~\cite{ref:conf/popl/Cousot97}, security analysis~\cite{ref:conf/popl/GiacobazziM04}, and many more~\cite{ref:conf/csl/CousotC14}. There are numerous general static analyzers for popular programming languages leveraging abstract interpretation, such as Infer for Java and C/C++~\cite{ref:conf/nfm/CalcagnoD11}, \mbox{Frama-C} for C~\cite{ref:conf/sefm/CuoqKKPSY12}, and Soot for Java~\cite{ref:conf/issta/Sherman18}. However, there are only few tools using abstract interpretation in the data science domain.

\Citeauthor*{ref:conf/icse/SuboticMS22}~\cite{ref:conf/icse/SuboticMS22} propose an abstract interpretation framework for data science notebooks to create a dependency graph of notebook cells by inferring preconditions on the cells. They extend this framework by a data leakage static analysis for machine learning notebooks~\cite{ref:conf/tase/DrobnjakovicSU24}. \Citeauthor*{ref:conf/esop/UrbanM18}~\cite{ref:conf/esop/UrbanM18} use abstract interpretation to automatically detect unused input data in data science programs, and extend this approach to infer expectations on input data~\cite{ref:series/isrl/Urban23}. However, these approaches focus on general properties of data science programs and do not consider the shape and size of the used data.

\Citeauthor*{ref:10.1145/3689609.3689996}~\cite{ref:10.1145/3689609.3689996} provide a high-level linter for data science programs by inferring abstract data types using abstract interpretation. However, they do not consider the shape and size of involved data. \Citeauthor*{ref:conf/pldi/NegriniSU23}~\cite{ref:conf/pldi/NegriniSU23} leverage abstract interpretation to create a control flow graph of data transformations in Python notebooks. They use this graph to infer the column names of data frames in Python notebooks. However, this is a first proof of concept for data frames in Python and does not consider unnamed columns or the number of columns and rows of data frames. \Citeauthor*{ref:journals/access/ZhuangL22}~\cite{ref:journals/access/ZhuangL22} present a tool to perform type checking on the columns of data frames in Python using abstract interpretation. However, the tool focuses on the \emph{pandas} library for Python and neither considers the size nor index-based accesses of data frames. Additionally, we find that none of the aforementioned approaches analyzing data frames evaluate their approach on real-world code.

In the context of~R, \emph{lintr}~\cite{ref:journals/jossw/HesterACHKPR25} is a linter for R~code that mostly relies on XPath expressions to identify possible syntax errors and semantic issues. It is limited to the syntactical perspective of an R~program and does not take data flow or control flow information into account. Similarly, \emph{rstatic}~\cite{ref:phd/us/Ulle21} is a static analysis tool for~R to search for syntax patterns, execution paths, and data propagation, which is extended by a theoretical type inference strategy for~R. However, \emph{rstatic} is limited to constant propagation and basic patterns, and does not use fixpoint iterations at loops nor considers modifications or transformations of the data. In contrast, \flowR~\cite{ref:10.1145/3763087} is a static data flow analyzer for~R providing a data flow and control flow graph. Besides value resolution via constant propagation, \flowR does not provide a mechanism to infer the shape and size of data in R~programs. To our knowledge, there is no existing approach that infers the shape of data frames in~R or that is evaluated on real-world code.

\section{Motivating Example}\label{sec:motivation}

Usually, when data frames are used in data science scripts, certain assumptions are made about the shape of the data frames, such as the existence of a column with a specific name or the number of columns and rows in the data frame. This is particularly the case when a column or row of a data frame is accessed by name or index. However, different transformations on data frames may affect the shape of the involved data frames.

Consider the example R~code in \cref{code:motivating_example}, that creates a data frame~\code{data} with the columns "id", "age", and "score", performs several transformations on the data frame using the pipe operator~\code{\%>\%}, and prints the mean of the column "score" to the console:

\begin{minted}[breaklines=false,escapeinside=||,caption={Example R~code manipulating data frames.},captionpos=b,label={code:motivating_example}]{R}
library(dplyr)

|\label{line:data}|data <- data.frame(
  id = c(1, 2, 3, 4),
  age = c(25, 50, 32, 28),
  score = c(65, 85, 40, 90)
)

data <- data %>%
|\label{line:filter}|  filter(age > 20) %>%
|\label{line:mutate}|  mutate(level = score / max(score)) %>%
|\label{line:select}|  select(id, age, level)

|\label{line:print}|print(mean(data$score))
\end{minted}

Although the operations in this example are straightforward, \lref{line:print} contains a subtle error. While accessing the column "score" to print the average score on the data frame \code{data} looks fine, the \code{select} operation in \lref{line:select} removes the column "score" from \code{data}. Hence, the column "score" does not exist in the data frame \code{data} in \lref{line:print} resulting in an error at runtime. While it might be trivial to identify the error in the small example above, this gets much harder in complex real-world R~scripts.

When tracking the shape of the data frame in the above example using abstract interpretation, we can infer the possible column names of the data frame \code{data} at the \code{print} statement in \lref{line:print}. With this information, we are able to detect that the accessed column "score" cannot exist in the data frame \code{data} in \lref{line:print}, as it has been removed by the preceding \code{select} operation.

\section{Approach}\label{sec:approach}

To address \ref{rq1}, we propose an abstract interpretation approach to infer the shape of data frames in R~programs. For this approach, we first analyze the most common operations on data frames in real-world R~scripts. Based on these results, we define an abstract domain for the shape of data frames and abstract the identified concrete data frame operations to abstract data frame operations. We then define the abstract semantics of these abstract operations with respect to our data frame shape domain.

\begin{table}
\newcounter{rowscount}\def\placecount{%
    \leavevmode\stepcounter{rowscount}%
    \llap{%
        \BeginAccSupp{method=escape,ActualText={}}%
        \tiny\sffamily\color{gray}\therowscount\ \ %
        \EndAccSupp{}%
    }%
}
\caption{The \num{20}~most used data frame-related functions in R~scripts from publications (based on \num{\ConceptEvalScripts}~scripts)~\cite{ref:conf/msr/SihlerPSTDD24}.}
\fontsize{8}{9.6}\selectfont
\begin{minipage}[b]{0.48\linewidth}
\begin{tabularx}{\linewidth}{>{\placecount\ttfamily}X @{\hspace{5pt}} r @{\hspace{5pt}} r}
\toprule
\multicolumn{1}{l}{function name} & of calls & of files\\
\midrule
data.frame     & \qty{0.98}{\percent}  & \qty{38}{\percent}\\
subset         & \qty{0.97}{\percent}  & \qty{22}{\percent}\\
colnames       & \qty{0.96}{\percent}  & \qty{32}{\percent}\\
nrow           & \qty{0.88}{\percent}  & \qty{30}{\percent}\\
names          & \qty{0.83}{\percent}  & \qty{30}{\percent}\\
rownames       & \qty{0.77}{\percent}  & \qty{18}{\percent}\\
cbind          & \qty{0.76}{\percent}  & \qty{30}{\percent}\\
as.data.frame  & \qty{0.67}{\percent}  & \qty{26}{\percent}\\
filter         & \qty{0.67}{\percent}  & \qty{17}{\percent}\\
read.csv       & \qty{0.66}{\percent}  & \qty{48}{\percent}\\
\bottomrule
\end{tabularx}
\end{minipage}
\hfill
\begin{minipage}[b]{0.48\linewidth}
\smash{\rlap{\raisebox{1pt}{\tiny\sffamily \faCaretDown~continued from the left}}}\endgraf%
\begin{tabularx}{\linewidth}{>{\placecount\ttfamily}X @{\hspace{5pt}} r @{\hspace{5pt}} r}
\toprule
\multicolumn{1}{l}{function name} & of calls & of files\\
\midrule
mutate         & \qty{0.62}{\percent}  & \qty{20}{\percent}\\
select         & \qty{0.47}{\percent}  & \qty{18}{\percent}\\
rbind          & \qty{0.43}{\percent}  & \qty{36}{\percent}\\
group\_by      & \qty{0.34}{\percent}  & \qty{21}{\percent}\\
dim            & \qty{0.34}{\percent}  & \qty{17}{\percent}\\
write.csv      & \qty{0.26}{\percent}  & \qty{20}{\percent}\\
ncol           & \qty{0.25}{\percent}  & \qty{19}{\percent}\\
merge          & \qty{0.25}{\percent}  & \qty{18}{\percent}\\
head           & \qty{0.24}{\percent}  & \qty{19}{\percent}\\
summarize      & \qty{0.23}{\percent}  & \qty{20}{\percent}\\
\bottomrule
\end{tabularx}
\end{minipage}
\label{tab:dataframe_functions}
\end{table}

\paragraph{Common Data Frame Operations}\label{par:common_operations}
We extract the most used functions related to data frames from \num{\ConceptEvalScripts}~real-world R~scripts based on a dataset by \citeauthor{ref:conf/msr/SihlerPSTDD24}~\cite{ref:conf/msr/SihlerPSTDD24} to identify common data frame operations. For this, we manually collect \num{\ConceptEvalDfFunctions}~functions related to data frames from the R~documentation and packages such as \emph{dplyr}, a library for data manipulation in~R. We then filter the most common function calls of the R~scripts by these data frame functions. However, with R's~dynamic type system, many of these functions are polymorphic and also work with other data structures like lists and matrices. Hence, not all identified function calls relate to data frames. \Cref{tab:dataframe_functions} shows the \num{20}~most commonly used data frame functions by their static call-based and file-based usage frequency. Almost every second R~script~(\qty{\ConceptEvalReadCsvFilePercent}{\percent}) uses the \code{read.csv} function to load a data frame from an external file. In addition, \qty{\ConceptEvalDataFrameFilePercent}{\percent}~of the scripts create a data frame using the \code{data.frame} function, which itself makes up almost \qty{1}{\percent}~of all function calls over all scripts, underlining the omnipresence of data frames in R~scripts.

We categorize these most common operations on data frames into data frame
\begin{andlist}
    \item creation functions~(\code{data.frame}, \code{read.csv},~\dots)
    \item column name and size functions~(\code{colnames}, \code{nrow},~\dots)
    \item transformation functions~(\code{subset}, \code{cbind},~\dots)
\end{andlist}.
Additionally, based on the dataset by \citeauthor*{ref:conf/msr/SihlerPSTDD24}, we find that columns are predominantly accessed by constant name~(\qty{\ConceptEvalColNameAccessPercent}{\percent}), such as via \code{df\$colname} or \code{df[["colname"]]}, rather than by constant index~(\qty{\ConceptEvalColIndexAccessPercent}{\percent}), while row access is mostly done by constant index (\qty{\ConceptEvalRowIndexAccessPercent}{\percent}), such as via \code{df[1:10,~]}, rather than by constant name~(\qty{\ConceptEvalRowNameAccessPercent}{\percent}).

\begin{figure*}
\normalsize
\centering
\begin{tikzpicture}
    \matrix[column sep=6mm,matrix of math nodes,ampersand replacement=\&,nodes={inner sep=0pt}] (mat1) {%
        \begin{gathered}
            \code{data.frame}(\sigma_1 = \vec{x}_1,\ldots,\sigma_n = \vec{x}_n)\ \mapsto\ \code{create}([\sigma_1,\ldots,\sigma_n], k)\\
                \qquad\qquad\qquad\qquad\text{with}\ k = \max(\text{len}(\vec{x}_1),\ldots,\text{len}(\vec{x}_n))\\
            \begin{aligned}
            \code{filter}(\df,b)\ \mapsto\ &\code{filterRows}(\df,b)\\
            \code{mutate}(\df,\sigma_1 = \code{\_},\ldots,\sigma_n = \code{\_})\ \mapsto\ &\code{mutateCols}(\df,[\sigma_1,\ldots,\sigma_n])\\
            \code{select}(\df,\sigma_1,\ldots,\sigma_n)\ \mapsto\ &\code{subsetCols}(\df,[\sigma_1,\ldots,\sigma_n])
            \end{aligned}
        \end{gathered} \&
        \begin{aligned}
            \df\code{\$}\,\sigma\ \mapsto\ &\code{accessCols}(\df,[\sigma])\\
            \df\code{[}\vec{\sigma}\code{]}\ \mapsto\ &\code{accessCols}(\df,[\sigma_1,\ldots,\sigma_n])\\
            \df\code{[}\vec{i}\code{]}\ \mapsto\ &\code{accessCols}(\df,[i_1,\ldots,i_n])\\
            \df\code{[}\vec{i},\code{\_]}\ \mapsto\ &\code{accessRows}(\df,[i_1,\ldots,i_n])
        \end{aligned}\\
    };
\end{tikzpicture}
\caption{Simplified mapping of a subset of concrete data frame operations to abstract data frame operations.}
\label{fig:abstraction_mapping}
\end{figure*}

\paragraph{Data Frame Shape Domain}\label{par:data_frame_domain}
As data frames are mostly accessed and modified by column name and column index, and the rows of data frames are mostly accessed and assigned by index, we focus on the column names, number of columns, and number of rows for the shape of data frames. Therefore, we define the abstract domain~\DF to capture the shape of data frames as a product domain, consisting of an abstract domain for the column names~\N, number of columns~\C, and number of rows~\R:
\begin{equation*}
    \DF := \N \times \C \times \R
\end{equation*}
For the complete lattice of the product domain, bottom, top, the partial order, join, meet, and widening are defined component-wise based on the sub abstract domains~\N, \C, and~\R defined below.

We represent the column names~\N of data frames through what we call a (string) set range domain with a minimum set~\(A\) of column names that must exist in the data frame and a maximum set~\(B\) of column names that can possibly exist in the data frame, denoted as a set range pair~\setrange{A, B}:
\begin{equation*}
    \N := \{\setrange{A, B} \mid A \in \powerset{\strings}, B \in \powerset{\strings} \cup \{\topN\}, A \subseteq B\} \cup \{\bot\}
\end{equation*}
Since the number of elements in the minimum and maximum set have to be bounded in practice, the maximum set can also be~\topN, representing any column names. In the following, we assume~\topN to be the set of all strings~\strings, so that we can use normal set operations on~\topN. For the set range domain, top is given as the set range~\setrange{\emptyset, \topN} with an empty minimum set~\(\emptyset\) and~\topN as maximum set, and bottom is defined explicitly as \bot~symbol. The partial order, join, meet, and widening are defined analogously to an interval domain~\cite{ref:conf/popl/CousotC77,ref:journals/ftpl/Mine17}:
\begin{align*}
    \setrange{A_1, B_1} \lesseq \setrange{A_2, B_2} &:\Leftrightarrow (A_2 \subseteq A_1) \wedge (B_1 \subseteq B_2)\\
    \setrange{A_1, B_1} \join \setrange{A_2, B_2} &:= \setrange{A_1 \cap A_2, B_1 \cup B_2}\\
    \setrange{A_1, B_1} \meet \setrange{A_2, B_2} &:=
    \begin{cases}
        \setrange{A_1 \cup A_2, B_1 \cap B_2} & \text{if}\ A_1 \cup A_2\subseteq B_1 \cap B_2\\
        \bot & \text{otherwise}
    \end{cases}\\
    \setrange{A_1, B_1} \widen{} \setrange{A_2, B_2} &:= \left(
    \begin{cases}
        A_1 & \text{if}\ A_1 \subseteq A_2\\
        \emptyset & \text{otherwise}
    \end{cases},
    \begin{cases}
        B_1 & \text{if}\ B_2 \subseteq B_1\\
        \topN & \text{otherwise}
    \end{cases}
    \right)
\end{align*}
The partial order of the set range domain is defined by whether the first set range is included in the second set range. The join is defined as the intersection of the minimum sets and union of the maximum sets, while the meet is defined as the union of the minimum sets and intersection of the maximum sets if this results in a valid set range and~\bot, otherwise. Widening is defined as setting the minimum set to~\(\emptyset\) if the minimum set decreases, and setting the maximum set to~\topN if the maximum set increases. Bottom~(\bot) is omitted here, since it is the least element of the partial order, neutral element of the join and widening, and absorbing element of the meet. In practice, we only store the minimum set and all elements of the maximum set that are not in the minimum set, as the minimum set always has to be a subset of the maximum set.

For the set range domain, the concretization~\(\concretization: \N \rightarrow \mathcal{N}\) and abstraction function~\(\abstraction: \mathcal{N} \rightarrow \N\) are defined as follows, forming a Galois connection~\((\mathcal{N}, \subseteq) \galois{\concretization}{\abstraction} (\N, \lesseq)\) with \(\mathcal{N} := \powerset{\powerset{\strings}}\):
\begin{align*}
    \concretization(\bot) &:= \emptyset\\
    \concretization(\setrange{A, B}) &:= \{\, x \in \powerset{\strings} \mid A \subseteq x, x \subseteq B \,\}\\
    \abstraction(X) &:=
    \begin{cases}
        \bot & \text{if}\ X = \emptyset\\
        \setrange{\bigcap_{x \in X} x, \bigcup_{x \in X} x} & \text{otherwise}
    \end{cases}
\end{align*}

For example, in \lref{line:data} of the example in \cref{code:motivating_example}, we infer the set range \setrange{\{\text{"id"}, \text{"age"}, \text{"score"}\}, \{\text{"id"}, \text{"age"}, \text{"score"}\}} for the column names of the data frame \code{data}, as we know that the columns "id", "age", and "score" must exist in \code{data} and there are no other possible column names in \code{data} resulting from \code{data.frame}.

As observed in \hyperref[par:common_operations]{\textit{Common Operations}} above, data transformations, such as \code{filter}, \code{group\_by}, and \code{merge}, whose resulting data frame shape depends on the values inside the data frame operand, are very common in real-world R~scripts. Hence, the number of columns and rows must be over-approximated in many cases. Therefore, we use a (closed) positive interval domain for the number of columns~\C and number of rows~\R with a positive lower bound~\(a\) and a positive, possibly infinite upper bound~\(b\):
\begin{equation*}
    \C = \R := \{\, [a, b] \mid a \in \naturals, b \in \naturals \cup \{+\infty\}, a \leq b \,\} \cup \{\bot\}
\end{equation*}
For the positive interval domain, top is given as the interval~\([0, +\infty]\) and bottom is defined explicitly as the \bot~symbol to avoid multiple representations of bottom by empty intervals. The partial order, join, meet, and widening are defined according to the conventional interval domain~\cite{ref:conf/popl/CousotC77,ref:journals/ftpl/Mine17}:
\begin{align*}
    [a_1, b_1] \lesseq [a_2, b_2] &:\Leftrightarrow (a_2 \leq a_1) \wedge (b_1 \leq b_2)\\
    [a_1, b_1] \join [a_2, b_2] &:= [\min(a_1, a_2), \max(b_1, b_2)]\\
    [a_1, b_1] \meet [a_2, b_2] &:=
    \begin{cases}
        [a_\textit{max}, b_\textit{min}] & \text{if}\ a_\textit{max} \leq b_\textit{min}\\
        \bot & \text{otherwise}
    \end{cases}\\
    \quad \text{where}\ a_\textit{max} &= \max(a_1, a_2)\ \text{and}\ b_\textit{min} = \min(b_1, b_2)\\
    [a_1, b_1] \widen{} [a_2, b_2] &:= \left[
    \begin{cases}
        a_1 & \text{if}\ a_1 \leq a_2\\
        0 & \text{otherwise}
    \end{cases},
    \begin{cases}
        b_1 & \text{if}\ b_1 \geq b_2\\
        +\infty & \text{otherwise}
    \end{cases}
    \right]
\end{align*}

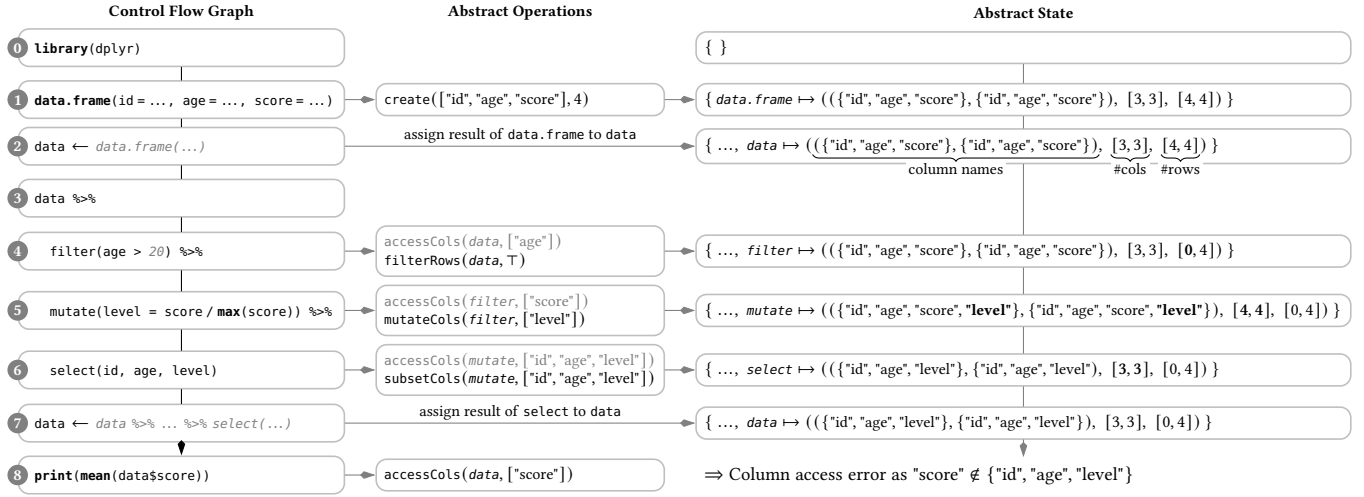
\begin{figure*}
    \centering
    \begingroup%
\xlstsetmintedstyle{plain}%
\newcommand*{\bracemarker}[2]{\marker{#1-1}{#2}\marker{#1-2}}%
\begin{tikzpicture}[%
        line cap=round,line join=round,font=\scriptsize,
        title/.style={font=\scriptsize\bfseries},
        vertex/.style={draw=lightgray,semithick,rounded corners=4pt,align=left,text width=4.1cm},
        operation/.style={draw=lightgray,semithick,rounded corners=4pt,align=left,text width=3.6cm},
        state/.style={draw=lightgray,semithick,rounded corners=4pt,align=left,text width=8.45cm},
        vertexid/.style={circle,white,fill=gray,align=center,inner sep=1pt,execute at begin node=\vphantom{012345689},font=\scriptsize\bfseries}
    ]%
    \matrix[column sep=4mm,row sep=1mm] {
        \node[title] {Control Flow Graph}; & \node[title] {Abstract Operations}; & \node[title] {Abstract State};\\[-1mm]
        \node[vertex] (library) {\begin{minted}[xleftmargin=3pt,aboveskip=-1pt,belowskip=0pt,breaklines=false,escapeinside=||]{R}
library(dplyr)
        \end{minted}}; & &
        \node[state] (state0) {
            \(\{\,\,\}\)
        };\\[2pt]
        \node[vertex] (dataframe) {\begin{minted}[xleftmargin=3pt,aboveskip=-1pt,belowskip=0pt,breaklines=false,escapeinside=||]{R}
data.frame(id|\;|=|\;\codedots|, age|\;|=|\;\codedots|, score|\;|=|\;\codedots|)
        \end{minted}}; &
        \node[operation] (create) {
            \(\code{create}([\text{"id"}, \text{"age"}, \text{"score"}], 4)\)
        }; &
        \node[state] (state1) {
            \(\{\,\code{\itshape data.frame} \mapsto (\setrange{\{\text{"id"}, \text{"age"}, \text{"score"}\}, \{\text{"id"}, \text{"age"}, \text{"score"}\}},\ [3, 3],\ [4, 4])\,\}\)
        };\\
        \node[vertex] (assignment1) {\begin{minted}[xleftmargin=3pt,aboveskip=-1pt,belowskip=0pt,breaklines=false,escapeinside=||]{R}
data <- |\textcolor{gray}{\itshape data.frame(\codedots)}|
        \end{minted}}; & &
        \node[state] (state2) {
            \(\{\,\codedots,\ \code{\itshape data} \mapsto (\bracemarker{colnames}{\setrange{\{\text{"id"}, \text{"age"}, \text{"score"}\}, \{\text{"id"}, \text{"age"}, \text{"score"}\}}},\ \bracemarker{cols}{[3, 3]},\ \bracemarker{rows}{[4, 4]})\,\}\)
        };\\[2pt]
        \node[vertex] (pipe) {\begin{minted}[xleftmargin=3pt,aboveskip=-1pt,belowskip=0pt,breaklines=false,escapeinside=||]{R}
data %>%
        \end{minted}}; & & \\
        \node[vertex] (filter) {\begin{minted}[xleftmargin=3pt,aboveskip=-1pt,belowskip=0pt,breaklines=false,escapeinside=||]{R}
  filter(age > 20) %>%
        \end{minted}}; &
        \node[operation] (filterRows) {
            \textcolor{gray}{\(\code{accessCols}(\code{\itshape data}, [\text{"age"}])\)}\\
            \(\code{filterRows}(\code{\itshape data}, \top)\)
        }; &
        \node[state] (state3) {
            \(\{\,\codedots,\ \code{\itshape filter} \mapsto (\setrange{\{\text{"id"}, \text{"age"}, \text{"score"}\}, \{\text{"id"}, \text{"age"}, \text{"score"}\}},\ [3, 3],\ [\mathbf{0}, 4])\,\}\)
        };\\
        \node[vertex] (mutate) {\begin{minted}[xleftmargin=3pt,aboveskip=-1pt,belowskip=0pt,breaklines=false,escapeinside=||]{R}
  mutate(level = score|\;|/|\;|max(score)) %>%
        \end{minted}}; &
        \node[operation] (mutateCols) {
            \textcolor{gray}{\(\code{accessCols}(\code{\itshape filter}, [\text{"score"}])\)}\\
            \(\code{mutateCols}(\code{\itshape filter}, [\text{"level"}])\)
        }; &
        \node[state] (state4) {
            \(\{\,\codedots,\ \code{\itshape mutate} \mapsto (\setrange{\{\text{"id"}, \text{"age"}, \text{"score"}, \textbf{"level"}\}, \{\text{"id"}, \text{"age"}, \text{"score"}, \textbf{"level"}\}},\ [\mathbf{4}, \mathbf{4}],\ [0, 4])\,\}\)
        };\\
        \node[vertex] (select) {\begin{minted}[xleftmargin=3pt,aboveskip=-1pt,belowskip=0pt,breaklines=false,escapeinside=||]{R}
  select(id, age, level)
        \end{minted}}; &
        \node[operation] (subsetCols) {
            \textcolor{gray}{\(\code{accessCols}(\code{\itshape mutate}, [\text{"id"}, \text{"age"}, \text{"level"}])\)}\\
            \(\code{subsetCols}(\code{\itshape mutate}, [\text{"id"}, \text{"age"}, \text{"level"}])\)
        }; &
        \node[state] (state5) {
            \(\{\,\codedots,\ \code{\itshape select} \mapsto (\setrange{\{\text{"id"}, \text{"age"}, \text{"level"}\}, \{\text{"id"}, \text{"age"}, \text{"level"}},\ [\mathbf{3}, \mathbf{3}],\ [0, 4])\,\}\)
        };\\
        \node[vertex] (assignment2) {\begin{minted}[xleftmargin=3pt,aboveskip=-1pt,belowskip=0pt,breaklines=false,escapeinside=||]{R}
data <- |\textcolor{gray}{\itshape data \%>\% \codedots\ \%>\% select(\codedots)}|
        \end{minted}}; & &
        \node[state] (state6) {
            \(\{\,\codedots,\ \code{\itshape data} \mapsto (\setrange{\{\text{"id"}, \text{"age"}, \text{"level"}\}, \{\text{"id"}, \text{"age"}, \text{"level"}\}},\ [3, 3],\ [0, 4])\,\}\)
        };\\[2pt]
        \node[vertex] (print) {\begin{minted}[xleftmargin=3pt,aboveskip=-1pt,belowskip=0pt,breaklines=false,escapeinside=||]{R}
print(mean(data$score))
        \end{minted}}; &
        \node[operation] (access) {
            \(\code{accessCols}(\code{\itshape data}, [\text{"score"}])\)
        }; &
        \node[state,draw=none] (state7) {\fontsize{7}{8}\selectfont\(\Rightarrow\) Column access error as \(\text{"score"} \notin \{\text{"id"}, \text{"age"}, \text{"level"}\}\)};\\
    };
    \node[vertexid] at (library.west) {\hypertarget{cfg:vertex0}{0}};
    \node[vertexid] at (dataframe.west) {\hypertarget{cfg:vertex1}{1}};
    \node[vertexid] at (assignment1.west) {\hypertarget{cfg:vertex2}{2}};
    \node[vertexid] at (pipe.west) {\hypertarget{cfg:vertex3}{3}};
    \node[vertexid] at (filter.west) {\hypertarget{cfg:vertex4}{4}};
    \node[vertexid] at (mutate.west) {\hypertarget{cfg:vertex5}{5}};
    \node[vertexid] at (select.west) {\hypertarget{cfg:vertex6}{6}};
    \node[vertexid] at (assignment2.west) {\hypertarget{cfg:vertex7}{7}};
    \node[vertexid] at (print.west) {\hypertarget{cfg:vertex8}{8}};
    \draw[] (library) -- (dataframe);
    \draw[] (dataframe) -- (assignment1);
    \draw[] (assignment1) -- (pipe);
    \draw[] (pipe) -- (filter);
    \draw[] (filter) -- (mutate);
    \draw[] (mutate) -- (select);
    \draw[] (select) -- (assignment2);
    \draw[-Kite] (assignment2) -- (print);
    \draw[-Kite,gray] (dataframe) -- (create);
    \draw[-Kite,gray] (create) -- (state1);
    \draw[-Kite,gray] (assignment1) -- node[midway,above=-2pt,black] {assign result of \code{data.frame} to \code{data}} (state2);
    \draw[-Kite,gray] (filter) -- (filterRows);
    \draw[-Kite,gray] (filterRows) -- (state3);
    \draw[-Kite,gray] (mutate) -- (mutateCols);
    \draw[-Kite,gray] (mutateCols) -- (state4);
    \draw[-Kite,gray] (select) -- (subsetCols);
    \draw[-Kite,gray] (subsetCols) -- (state5);
    \draw[-Kite,gray] (assignment2) -- node[midway,above=-2pt,black] {assign result of \code{select} to \code{data}} (state6);
    \draw[-Kite,gray] (print) -- (access);
    \draw[gray] (state0) -- (state1);
    \draw[gray] (state1) -- (state2);
    \draw[gray] (state2) -- (state3);
    \draw[gray] (state3) -- (state4);
    \draw[gray] (state4) -- (state5);
    \draw[gray] (state5) -- (state6);
    \draw[-Kite,gray] (state6) -- (state7);
\end{tikzpicture}%
\begin{tikzpicture}[overlay,remember picture,gray,brace/.style={decoration={brace,mirror,raise=-0.5pt},decorate,line cap=round,line join=round,black,font=\scriptsize}]
    \draw[brace] (colnames-1.south) to[edge node={node[below=2pt,fill=white,inner ysep=0.5pt,inner xsep=1pt] {\vphantom{l}column names}}] (colnames-2.south);
    \draw[brace] (cols-1.south) to[edge node={node[below=2pt,fill=white,inner ysep=0.5pt,inner xsep=1pt] {\vphantom{l}\#cols}}] (cols-2.south);
    \draw[brace] (rows-1.south) to[edge node={node[below=2pt,fill=white,inner ysep=0.5pt,inner xsep=1pt] {\vphantom{l}\#rows}}] (rows-2.south);
\end{tikzpicture}%
\endgroup%
    \caption{The concept of abstracting the concrete data frame operations to abstract operations and applying their abstract semantics with respect to the data frame shape domain for the example in \cref{code:motivating_example}.}
    \label{fig:shape_inference_concept}
\end{figure*}

The concretization~\concretization and abstraction function~\abstraction of the positive interval domain are defined as follows, creating a Galois connection~\((\mathcal{C}, \subseteq) \galois{\concretization}{\abstraction} (\C, \lesseq)\) and \((\mathcal{R}, \subseteq) \galois{\concretization}{\abstraction} (\R, \lesseq)\) with \(\mathcal{C} = \mathcal{R} := \powerset{\naturals}\):
\begin{align*}
    \concretization(\bot) &:= \emptyset\\
    \concretization([a, b]) &:= \{\, x \in \naturals \mid a \leq x \leq b \,\}\\
    \abstraction(X) &:=
    \begin{cases}
        \bot & \text{if}\ X = \emptyset\\
        [\min(X), \max(X)] & \text{otherwise}
    \end{cases}
\end{align*}

For example, for the variable \code{data} in \lref{line:data} of the example in \cref{code:motivating_example}, we infer the interval~\([3, 3]\) for the number of columns and the interval~\([4, 4]\) for the number of rows, as the data frame \code{data} is created with exactly \(3\)~columns and \(4\)~rows via \code{data.frame}. Hence, for the overall data frame shape of the variable \code{data}, we get the tuple~\((\setrange{\{\text{"id"}, \text{"age"}, \text{"score"}\}, \{\text{"id"}, \text{"age"}, \text{"score"}\}},\ [3, 3],\ [4, 4])\) in the data frame shape domain~\DF with the set range for the column names and intervals for the number of columns and rows.

Until now, we defined the data frame shape domain as a product domain \(\DF := \N \times \C \times \R\) with independent abstract domains for the column names~\N and number of columns~\C. However, if we know which column names must exist and which can possibly exist in a data frame, we can refine the inferred number of columns by limiting them to the number of column names in the minimum and maximum set of~\N. Conversely, we can refine the maximum set of the column names if we know from the inferred number of columns that there cannot be additional possible column names besides those in the minimum set. Therefore, we refine the data frame shape domain~\DF as a partially reduced product domain using the reduction~\rhoDF defined as follows:
\begin{align*}
    \rho_1(\setrange{A, B},\ [a, b],\ \X[r]) &:= \begin{cases}
        \bot                               & \text{if}\ |A| > b\\
        (\setrange{A, A},\ [a, b],\ \X[r]) & \text{if}\ |A| = b\\
        (\setrange{A, B},\ [a, b],\ \X[r]) & \text{otherwise}
    \end{cases}\\
    \rho_2(\setrange{A, B},\ [a, b],\ \X[r]) &:= (\setrange{A, B},\ [a, b] \meet [|A|, |B|],\ \X[r])\\
    \rhoDF(\X[d\!f]) &:= \rho_2(\rho_1(\X[d\!f]))
\end{align*}
The reduction is split into two reduction steps. First, the inferred column names are reduced using~\(\rho_1\) by limiting the maximum set~\(B\) of the inferred column names to the minimum set~\(A\) if we know that the upper bound of the number of columns~\(b\) equals the number of column names in the minimum set~\(A\). Second, we refine the inferred column count by meeting the current number of columns~\([a, b]\) with the number of inferred column names~\([|A|, |B|]\). The reduction is applied after every product domain operation to refine the inferred data frame shape.

\newcommand*{\vertex}[1]{\smash{\raisebox{0.5pt}{\tikz[baseline=(text.base)]\node[circle,white,fill=gray,align=center,inner sep=1pt,execute at begin node=\vphantom{012345689},font=\scriptsize\bfseries] (text) {\hyperlink{cfg:vertex#1}{#1}};}}}

\paragraph{Abstract Operations}\label{par:abstract_operations}
Since there are many data frame operations from different libraries with similar semantics, such as \code{read.csv} from base~R and \code{read\_csv} from \emph{readr}, \code{subset} from base~R and \code{select} from \emph{dplyr}, and \code{transform} from base~R and \code{mutate} from \emph{dplyr}, we abstract these concrete data frame operations to sets of abstract data frame operations to summarize them by semantics.

\Cref{fig:abstraction_mapping} shows the mapping of the concrete data frame operations of the example in \cref{code:motivating_example} to abstract data frame operations. Here, \df~represents the data frame operand of each data frame operation, \(\sigma_1,\ldots,\sigma_n\) represent strings or symbol names, \(i_1,\ldots,i_n\) represent integers, \(b\) represents a Boolean value, and \(\vec{x}_1,\ldots,\vec{x}_n\) represent arbitrary vectors. However, these mappings are simplified for readability and omit the following properties:
\begin{andlist}
    \item the values of the arguments that are not data frames like strings and integers need to be resolved by other abstract domains and can therefore also be top
    \item transformation functions that implicitly access columns in their arguments, such as \code{filter}, \code{select}, and \code{mutate} are additionally mapped to \code{accessCols} for the implicitly accessed columns
    \item there are many edge cases for these concrete operations that are omitted here but are supported in our analysis
\end{andlist}.
We describe the mapping of the concrete data frame operations to abstract data frame operations based on our running code example in \cref{code:motivating_example}.

\Cref{fig:shape_inference_concept} shows the process of abstracting the concrete data frame operations of the example to abstract operations and applying their abstract semantics to the current abstract state. The left column of the figure shows a strongly simplified control flow graph of the code example in \cref{code:motivating_example}. In the control flow graph, the assignments themselves come after the right hand-side of the assignments in the order of evaluation. The middle column shows the mapped abstract data frame operations for each concrete data frame operation with the data frame operand and the arguments. In the right column, for every control flow graph vertex, the inferred abstract state that maps the data frame variables and expressions to their inferred abstract data frame shapes after applying the abstract semantics of the abstract operations is shown. In the abstract state, the~"\codedots" indicates that the mapping shown next is newly added to the already existing mappings in the abstract state.

\begin{figure*}
\centering
\begin{tikzpicture}
    \matrix[column sep=6mm,matrix of math nodes,ampersand replacement=\&,nodes={inner sep=0pt}] (mat1) {%
        \begin{aligned}
            \expr{n}{\code{create}([\sigma_1,\ldots,\sigma_n],k)}{\X[n]} &:= \setrange{\{\sigma_1,\ldots,\sigma_n\}, \{\sigma_1,\ldots,\sigma_n\}}\\
            \expr{c}{\code{create}([\sigma_1,\ldots,\sigma_n],k)}{\X[c]} &:= [n,n]\\
            \expr{r}{\code{create}([\sigma_1,\ldots,\sigma_n],k)}{\X[r]} &:= [k,k]
        \end{aligned} \&
        \begin{aligned}
            \expr{r}{\code{filterRows}(\code{FALSE})}{[a,b]} &:= [0,0]\\
            \expr{r}{\code{filterRows}(\code{TRUE})}{[a,b]} &:= [a,b]\\
            \expr{r}{\code{filterRows}(\top)}{[a,b]} &:= [0,b]
        \end{aligned}\\
    };
    \matrix[below=1mm of mat1,column sep=6mm,matrix of math nodes,ampersand replacement=\&,nodes={inner sep=0pt}] (mat2) {%
        \begin{aligned}
            \expr{n}{\code{mutateCols}([\sigma_1,\ldots,\sigma_n])}{\setrange{A, B}} &:= \setrange{A, B} \dotcup{} \{\sigma_1,\ldots,\sigma_n\}\\
            \expr{c}{\code{mutateCols}([\sigma_1,\ldots,\sigma_n])}{[a,b]} &:= [\max(a,n),b + n]
        \end{aligned} \&
        \begin{aligned}
            \expr{n}{\code{subsetCols}([\sigma_1,\ldots,\sigma_n])}{\setrange{A, B}} &:= \setrange{A, B} \dotcap{} \{\sigma_1,\ldots,\sigma_n\}\\
            \expr{c}{\code{subsetCols}([\sigma_1,\ldots,\sigma_n])}{[a,b]} &:= [\min(a,n),\min(b,n)]
        \end{aligned}\\
    };
\end{tikzpicture}
\caption{Abstract semantics of a few abstract data frame operations for the column names, column count, and row count.}
\label{fig:abstract_semantics}
\end{figure*}

The first data frame operation in the code example is the data frame creation function \code{data.frame} at control flow vertex~\vertex{1}. Using the mapping for \code{data.frame} in \Cref{fig:abstraction_mapping}, the function is mapped to the abstract data frame operation \(\code{create}([\text{"id"}, \text{"age"}, \text{"score"}], 4)\) with the names of the arguments "id", "age", and "score" as column names. R~implicitly recycles vectors of arguments of the \code{data.frame} function to match the length of the longest vector argument. Hence, the number of created rows~\(k\) is the length of the longest vector argument, which is~\(4\) in our example.

Subsequently, the function \code{filter} of \emph{dplyr} for filtering the rows of a data frame based on a condition at vertex~\vertex{4} is mapped to the abstract operation \(\code{filterRows}(\code{\itshape data}, \top)\) using the corresponding mapping in \Cref{fig:abstraction_mapping}. The data frame operand of the abstract operation is the variable \code{data}, which is implicitly given as first argument to \code{filter} via the pipe operator~\code{\%>\%}, and the filter condition~\(b\) is~\top in this case, as it depends on the values inside the data frame and cannot be resolved to a constant. Additionally, as the \code{filter} operation implicitly accesses the column "age" here, \code{filter} is also mapped to \code{accessCols} for the implicitly accessed column "age".

Using the mapping for \code{mutate} in \Cref{fig:abstraction_mapping}, the function \code{mutate} of \emph{dplyr} for creating or modifying columns of a data frame based on existing columns at vertex~\vertex{5} is mapped to the abstract operation \(\code{mutateCols}(\code{\itshape filter}, [\text{"level"}])\) with the result of the preceding \code{filter} function as data frame operand, which is again implicitly provided via the pipe operator. The names of the arguments of the function call are the created or modified columns, which is the column "level" in this case. As we are only interested in the resulting data frame shape, the values of the arguments are ignored, denoted by~"\code{\_}". The \code{mutate} function is again also mapped to \code{accessCols} for the implicitly accessed column "score".

The function \code{select} for subsetting the columns of a data frame at vertex~\vertex{6} is mapped to \(\code{subsetCols}(\code{\itshape mutate}, [\text{"id"}, \text{"age"}, \text{"level"}])\) using the corresponding mapping in \Cref{fig:abstraction_mapping}. The data frame operand is given as the result of the preceding \code{mutate} function via the pipe operator, and the subsetted columns are the arguments "id", "age", and "level". The function is again also mapped to \code{accessCols} for the implicitly accessed columns "id", "age", and "level".

Data frames in~R can be accessed by constant column name via~\code{\$}, by name or index via~\code{[} to access multiple columns or rows, or by name or index via~\code{[[} to access a single column or cell. As described in \Cref{fig:abstraction_mapping}, column accesses by constant name via~\code{\$} and column accesses by name or index via~\code{[} are mapped to \code{accessCols} with the accessed column names or column indices as arguments. As with other programming languages as well, rows can be accessed via the first argument in an access with two access arguments. Hence, such row accesses via~\code{[} are mapped to the abstract operation \code{accessRows} with the accessed row indices as arguments. Analogously, column and row accesses via~\code{[[} are mapped to \code{accessCols} and \code{accessRows} as well. In our example, the column access~\code{\$} at control flow vertex~\vertex{8} is mapped to \code{accessCols} with the variable \code{data} as data frame operand and the accessed column "score" as argument, resulting in the abstract operation \(\code{accessCols}(\code{\itshape data}, [\text{"score"}])\).

\paragraph{Abstract Semantics}\label{par:abstract_semantics}
After mapping the concrete data frame operations to abstract data frame operations, we define the abstract semantics of these abstract operations with respect to the data frame shape domain~\DF in \Cref{fig:abstract_semantics}. The abstract expression semantics of an abstract data frame operation \emph{expr} are defined by the evaluation function \expr{d\!f}{\textit{expr}}{: \DF \rightarrow \DF} that transforms the abstract data frame shape of the data frame operand of the abstract operation into the resulting abstract data frame shape. Hence, we omit the data frame operand in the abstract operations here, as the abstract data frame shape of the operand is given to the expression function as argument. As the data frame shape domain~\DF is a product domain, we define the abstract semantics for the column names~\N, number of columns~\C, and number of rows~\R separately of each other as \expr{n}{\textit{expr}}{: \N \rightarrow \N}, \expr{c}{\textit{expr}}{: \C \rightarrow \C}, and \expr{r}{\textit{expr}}{: \R \rightarrow \R}, respectively. The expression semantics of the data frame shape domain~\DF is therefore given as \(\expr{d\!f}{\textit{expr}}{(\X[n],\X[c],\X[r])} := (\expr{n}{\textit{expr}}{\X[n]}, \expr{c}{\textit{expr}}{\X[c]}, \expr{r}{\textit{expr}}{\X[r]})\). After every application of the abstract semantics, the reduction~\rhoDF of the reduced product domain~\DF is used to refine the inferred data frame shape. We use the notation \(\setrange{A, B} \dotcup{} \{\sigma_1,\ldots,\sigma_n\}\) for the union of \(\{\sigma_1,\ldots,\sigma_n\}\) with both \(A\) and~\(B\), and \(\setrange{A, B} \dotcap{} \{\sigma_1,\ldots,\sigma_n\}\) for the intersection of \(\{\sigma_1,\ldots,\sigma_n\}\) with both \(A\) and~\(B\).

However, we also simplify the abstract semantics of the abstract operations here and omit the following properties for brevity:
\begin{andlist}
    \item we only describe the semantics for the shape properties that are actually changed by the abstract operations and map the semantics for the other shape properties implicitly to the identity
    \item if an argument of an abstract operation like a subsetted column name is top, the respective shape property must be soundly over-approximated
    \item we omit the abstract semantics for shape properties that are bottom for simplicity
\end{andlist}.
We again use our running example shown in \Cref{fig:shape_inference_concept} to describe the abstract semantics of the abstract data frame operations in the following.

For the \code{create} operation at control flow vertex~\vertex{1}, using the abstract column names semantics for \code{create} defined in \Cref{fig:abstract_semantics}, we get the set range \(\setrange{\{\text{"id"}, \text{"age"}, \text{"score"}\}, \{\text{"id"}, \text{"age"}, \text{"score"}\}}\) for the column names with equal minimum and maximum set, as there cannot be any additional possible column names. Applying the abstract column count and row count semantics of the \code{create} operation, we get the interval \([3, 3]\) for the number of columns and \([4, 4]\) for the number of rows. This results in the abstract data frame shape \((\setrange{\{\text{"id"}, \text{"age"}, \text{"score"}\}, \{\text{"id"}, \text{"age"}, \text{"score"}\}},\ [3, 3],\ [4, 4])\) for the return value of the \code{data.frame} function in the abstract state. After evaluating the assignment at vertex~\vertex{2}, the variable \code{data} is mapped to the abstract value resulting from the \code{data.frame} function in the abstract state.

Applying the abstract semantics of the \code{filterRows} operation at vertex~\vertex{4}, we use the third definition of the abstract row count semantics for \code{filterRows} in \Cref{fig:abstract_semantics}, as the filter condition is~\top. Since the filter operation can remove an arbitrary number of rows but cannot add new rows, we soundly over-approximate the resulting number of rows by extending the lower bound down to~\(0\), resulting in the interval \([0, 4]\) for the number of rows for the return value of the \code{filter} function in the abstract state.

Using the abstract column names semantics for the \code{mutateCols} operation at vertex~\vertex{5} in \Cref{fig:abstract_semantics}, we create the union of the current column names with the mutated column "level", as the mutated column has to exist in the data frame after the mutation operation. Applying the abstract column count semantics of \code{mutateCols}, the number of mutated columns is added to the upper bound of the number of columns, as the mutated columns could be newly created, but could also have existed before, increasing the number of columns to~\([3, 4]\) for the return value of the \code{mutate} function in the abstract state. After applying the reduction~\rhoDF of the data frame shape domain~\DF, the number of columns is refined to~\([4, 4]\), as we know that the data frame has exactly \num{4}~column names.

For the \code{subsetCols} operation at vertex~\vertex{6}, applying the abstract column names semantics for \code{subsetCols} in \Cref{fig:abstract_semantics}, the current column names are intersected with the subsetted columns "id", "age", and "level". Using the abstract column count semantics for \code{subsetCols}, the resulting number of columns is the minimum of the lower and upper bound of the current number of columns and the number of subsetted columns. This results in the column names "id", "age", and "level" with the column "score" removed, and the interval~\([3, 3]\) for the number of columns for the return value of the \code{select} function in the abstract state. After evaluating the assignment at vertex~\vertex{7}, the variable \code{data} is mapped to the abstract data frame shape resulting from the \code{select} function in the abstract state. Therefore, at vertex~\vertex{8}, we observe that the accessed column "score" indeed cannot exist in the data frame \code{data}, as "score" is not in the set of possible column names \(\{\text{"id"}, \text{"age", \text{"level"}}\}\) of \code{data}.

\paragraph{Implementation}\label{par:implementation}
We provide a companion implementation of the data frame shape inference based on \flowR~\cite{ref:10.1145/3763087}, a static data flow analyzer for~R. \flowR generates a data flow and control flow graph of an R~program, handling most of R's dynamic features~\cite{ref:10.1145/3763087}. On top of \flowR, we perform abstract interpretation using a semantic-guided visitor on the control flow graph generated by \flowR for fixpoint iteration. Since many control flow constructs like \code{if}~conditions are expressions rather than statements in~R, which can also be overwritten, the control flow graph of \flowR consists of expressions without summarizing them to basic blocks.

The visitor forward-traverses the control flow graph of an R~program in depth-first order using a worklist algorithm and maps vertices representing data frame expressions to abstract data frame operations, as described in \Cref{fig:shape_inference_concept}. For the abstract operations, the abstract data frame shape of the data frame operand is resolved using the data flow graph of \flowR and our inferred abstract state at the currently visited vertex. All other arguments of the data frame operations, such as strings, integers, and vectors, are resolved to constants using constant propagation. Based on the abstract semantics defined in \cref{sec:approach}, the visitor applies the abstract semantics of the abstract data frame operations using the shape of the data frame operand to the abstract state at the currently visited vertex.

To ensure termination of the fixpoint iteration, we perform delayed widening at loop heads after a configurable number of visitations. However, we do not yet support interprocedural abstract interpretation for calls to user-defined functions using data frames. For data frames that are read from external files, such as CSV~files, we extract the column names using the CSV~header and read the number of columns and rows from the files if the files are locally available to retrieve the resulting data frame shape.

Currently, we support \num{41}~concrete data frame operations explicitly and map them to \num{22}~abstract data frame operations. This includes all \num{20}~most commonly used data frame functions listed in \Cref{tab:dataframe_functions}, except for \code{nrow}, \code{dim}, \code{ncol}, and \code{write.csv}, as they do not modify data frames. Additionally, there are \num{103}~other data frame operations that are not explicitly supported by our implementation, but that we handle as unknown data frame sources or transformations to at least infer that they return data frames to be able to refine the data frame shape in subsequent data frame operations.

Using the inferred data frame shapes, we provide a linter rule to validate name- and index-based column and row accesses of data frames in R~scripts. For this, we check at each identified data frame access operation whether the accessed column names, column indices, or row indices are satisfied by the inferred abstract data frame shape of the data frame operand of the access operation at the respective location.

\paragraph{\textbf{Summary of \ref{rq1}}}\label{par:rq1_summary}
We define a novel abstract domain for the shape of data frames as a partially reduced product domain consisting of sub abstract domains for the column names, number of columns, and number of rows. To summarize concrete data frame operations by their semantics, we map them to abstract data frame operations and define the abstract semantics of these abstract data frame operations with respect to the data frame shape domain. We provide an accompanying implementation of the data frame shape inference based on \flowR and a linter rule to detect invalid data frame accesses in R~programs using the inferred data frame shapes.

\section{Evaluation}\label{sec:evaluation}

For \ref{rq2}, to measure the correctness and effectiveness of our approach, we evaluate our analysis on a set of real-world R~scripts~\cite{ref:10.5281/zenodo.18927873} in two steps: First, we empirically verify the correctness of our approach on \num{\CorrEvalScripts}~executable real-world R~scripts based on ground truth from running the scripts. Second, we perform a large-scale evaluation on \num{\TotalEvalScripts}~real-world R~scripts to measure the number of inferred data frame shape constraints, identified data frame operations, detected invalid data frame accesses, and runtime.

\subsection{Correctness Evaluation}\label{sub:correctness}

Besides \num{\TestsTotal}~systematic unit tests with \qty{\TestsCoverage}{\percent}~coverage, we verify the correctness of our approach on \num{\CorrEvalScripts}~executable real-world R~scripts empirically by comparing the inferred data frame shapes with the actual data frame shapes when running the scripts. These R~scripts were manually reviewed and checked for executability and determinism~\cite{ref:10.5281/zenodo.18927873}, and all datasets read in the scripts are present, enabling our analysis to extract the data frame shape from the datasets.

To retrieve the ground truth, we leverage R's reflective capabilities to instrument the R~scripts to output at every variable definition whether the variable represents a data frame, as well as the column names, number of columns, and number of rows of the variable if it represents a data frame. We then use our analysis to statically infer the data frame shape for these variable definitions in the R~scripts. By comparing the actual data frame shapes with the inferred data frame shapes for these variable definitions, we determine how many data frames are identified by our analysis and how often the inferred shape is exact, an over-approximation, or an under-approximation of the actual shape.

Of the \num{\CorrEvalPoints}~variable definitions in the \num{\CorrEvalScripts}~executable R~scripts, \num{\CorrEvalDfPoints}~variable definitions~(\qty{\CorrEvalDfPointsPercent}{\percent}) represent data frames. We successfully inferred data frame shape constraints for \num{\CorrEvalCorrDf}~of the variable definitions that represent data frames~(\qty{\CorrEvalDfCorrDfPercent}{\percent}), over-approximating the remaining \num{\CorrEvalWrongEmpty}~undetected data frames implicitly by top~(\qty{\CorrEvalWrongEmptyPercent}{\percent}). There was only a single case where we inferred a data frame shape for a variable definition that does not represent a data frame due to a special edge case where we did not detect that the result of a row access was implicitly coerced to a vector.

By manually investigating the cases where we did not recognize data frames as such, we find that this is mostly due to current limitations of our implementation, such as
\begin{andlist}
    \item variables that are indirectly loaded into the environment from a binary file using \code{load}
    \item multiple possible method signatures for functions (e.g., for \code{aggregate})
    \item higher-order functions like \code{do.call} and \code{ddply}
    \item functions from domain-specific packages like \emph{astrochron} and \emph{inferference}
    \item data frames resulting from user-defined functions
\end{andlist}
which are not yet supported by the implementation.

\begin{figure*}%
    \centering%
    \captionsetup{captionskip=-2pt,nearskip=0pt,farskip=6pt}%
    \subfloat[]{%
        \includegraphics[width=\dimexpr\textwidth+5pt]{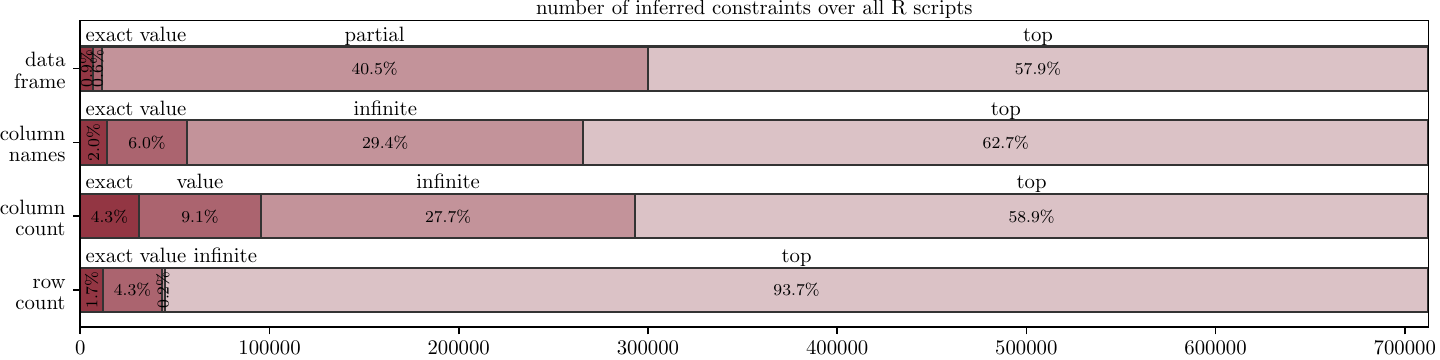}%
        \label{fig:inferred_shapes_all}%
    }\\%
    \captionsetup{captionskip=2pt,nearskip=0pt,farskip=6pt}%
    \subfloat[]{%
        \includegraphics[width=\textwidth]{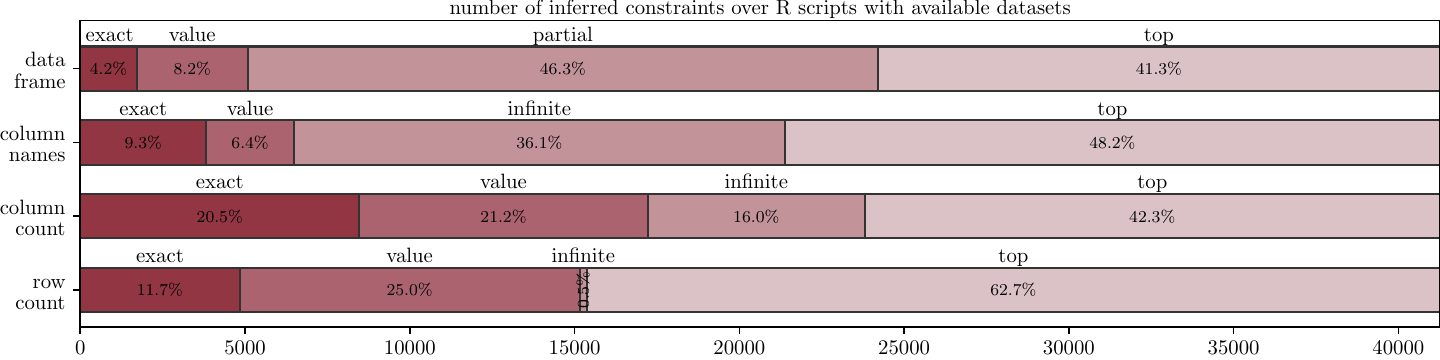}%
        \label{fig:inferred_shapes_datasets}%
    }%
    \caption{The total number of inferred data frame shapes over all analyzed R~scripts~(\cref{fig:inferred_shapes_all}) and R~scripts with all read datasets available~(\cref{fig:inferred_shapes_datasets}), categorized into top, partial information of infinite values, finite non-exact values, and exact values.}
    \label{fig:inferred_shapes}
\end{figure*}

Of the \num{\CorrEvalCorrDf} detected data frame variable definitions, we over-approximated the actual data frame shape in \num{\CorrEvalOverApprox}~cases~(\qty{\CorrEvalOverApproxPercent}{\percent}), where \num{\CorrEvalTop}~cases~(\qty{\CorrEvalTopPercent}{\percent}) were top, and inferred an exact data frame shape matching the actual shape in \num{\CorrEvalExact}~cases~(\qty{\CorrEvalExactPercent}{\percent}). There was not a single case where we under-approximated the actual shape, providing empirical evidence for the soundness of our approach. For the inferred data frame shapes with finite upper bounds~(neither~\topN nor~\(+\infty\)), using the mid-value of the inferred ranges, the number of inferred column names was, on average, \qty{\CorrEvalColNamesMidDiffTimes}{\({\times}\)}~the actual number of column names, the inferred column count was, on average, \qty{\CorrEvalColCountMidDiffTimes}{\({\times}\)}~the actual column count, and the inferred row count was, on average, \qty{\CorrEvalRowCountMidDiffTimes}{\({\times}\)}~the actual row count. While we are close to the actual value for the number of column names and the column count, the inferred row count is much greater than the actual row count, probably due to the fact that we have to over-approximate the row count for all filter and group operations.

\begin{finding}
Evaluating the correctness of our approach on ground truth extracted from \num{\CorrEvalScripts}~executable real-world R~scripts, we inferred data frame shape constraints for \qty{\CorrEvalDfCorrDfPercent}{\percent}~of the variables that actually represent data frames and empirically verify the soundness of our approach by never under-approximating the actual shape.
\end{finding}

\subsection{Effectiveness Evaluation}\label{sub:evaluation}

To determine the performance and effectiveness of our approach, we perform a large-scale evaluation on \num{\TotalEvalScripts} real-world R~scripts from publication artifacts~\cite{ref:10.5281/zenodo.18927873}. To identify all required datasets that are read in these scripts, we perform a pre-run of our analysis to get all paths to datasets that we would read to extract the data frame shape when analyzing the script. However, we can only automatically retrieve all read datasets for \num{\TotalEvalScriptsDatasets}~R~scripts~(\qty{\TotalEvalScriptsDatasetPercent}{\percent}) due to datasets that are not uploaded by the authors and invalid file paths, such as absolute file paths. Therefore, since many R~scripts load data frames from external files, as observed in \cref{sec:approach}, we have to over-approximate the read data frames for the majority of scripts for which we do not have the read datasets available.

\paragraph{Effectiveness}\label{par:effectiveness}
By running our analysis on all \num{\TotalEvalScripts}~real-world R~scripts, \num{\TotalEvalFailed}~scripts could not be analyzed due to our analysis running out of memory. Our analysis generally found data frames in \qty{\TotalEvalDfScriptsPercent}{\percent}~of the successfully analyzed R~scripts. On average, we mapped \(\num{\TotalEvalAvgNumOperations}\pm\num{\TotalEvalSdNumOperations}\)~data frame operations and inferred \(\num{\TotalEvalAvgConstraints}\pm\num{\TotalEvalSdConstraints}\)~data frame shapes for data frame operations per script, indicating large differences of the usage of data frames between the scripts.

We inferred data frame shapes for \num{\TotalEvalTotalConstraints}~data frame operations in total, accounting for \qty{\TotalEvalDfCallsPercent}{\percent}~of all function calls over all analyzed R~scripts, underlining the importance of analyzing data frames in~R. \Cref{fig:inferred_shapes_all} shows the number of inferred shape constraints overall, for the column names, column count, and row count, categorized into
\begin{andlist}
    \item \emph{top}
    \item \emph{partial}, where at least one shape property was not top
    \item \emph{infinite}, where the upper bound is infinite~(\topN or~\(+\infty\)), but the value is not top
    \item \emph{value}, where the upper bound is finite
    \item \emph{exact}, where the lower bound matches the upper bound
\end{andlist}.

\begin{figure*}%
    \centering%
    \captionsetup{captionskip=-12pt,nearskip=0pt,farskip=0pt}%
    \subfloat[]{%
        \includegraphics[width=0.245\textwidth]{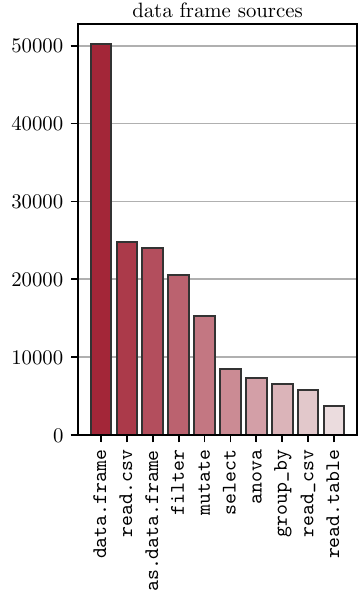}%
        \label{fig:dataframe_sources}%
    }%
    \hfill%
    \subfloat[]{%
        \includegraphics[width=0.245\textwidth]{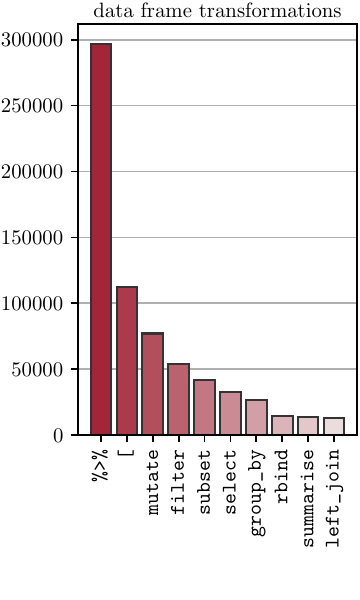}%
        \label{fig:dataframe_transformations}%
    }%
    \hfill%
    \subfloat[]{%
        \includegraphics[width=0.245\textwidth]{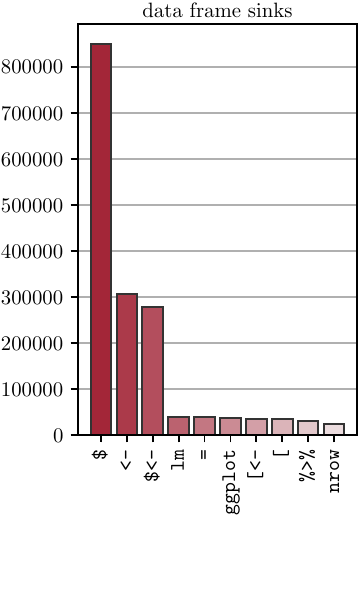}%
        \label{fig:dataframe_sinks}%
    }%
    \hfill%
    \subfloat[]{%
        \includegraphics[width=0.245\textwidth]{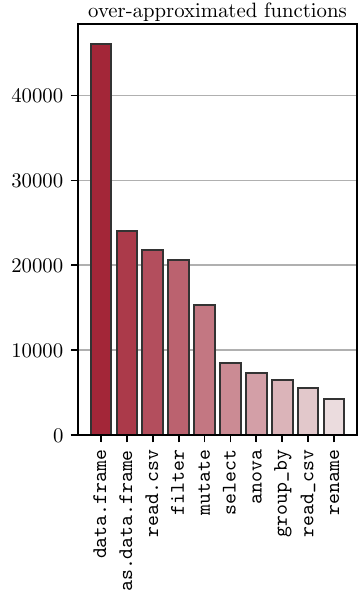}%
        \label{fig:overapproximated_functions}%
    }%
    \caption{The \num{10} most common data frame operations representing data frame sources~(\cref{fig:dataframe_sources}), data frame transformations~(\cref{fig:dataframe_transformations}), and data frame sinks~(\cref{fig:dataframe_sinks}). \cref{fig:overapproximated_functions} shows the 10 most common over-approximated data frame operations for which we inferred a data frame shape with infinite upper bounds for the returned data frame.}
    \label{fig:dataframe_functions}
\end{figure*}

Overall, for \qty{\TotalEvalDfShapeTopPercent}{\percent}~of the inferred data frame shapes, the column names, column count, and row count were top. However, this does not include implicit top values if we did not detect that an operation returns a data frame (which was the case for \qty{\CorrEvalWrongEmptyPercent}{\percent}~of the actual data frame operations in the correctness evaluation). For \qty{\TotalEvalDfShapePartialPercent}{\percent}~of the inferred data frame shapes, we inferred partial shape information where at least one shape property was not top. Subsequently, for \qty{\TotalEvalDfShapeValuePercent}{\percent}~of the inferred shapes, the column names, column count, and row count were not top, and, for~\qty{\TotalEvalDfShapeExactPercent}{\percent}, the column names, column count, and row count were exact, meaning that the inferred lower bound equals the inferred upper bound.

Regarding the individual shape properties, we inferred the most concrete constraints that are not top, finite values, and exact values for the column count of the data frames, followed by the column names, and inferred much fewer concrete constraints for the row count. However, the number of inferred constraints that are top, infinite, finite, or exact does not necessarily reflect the precision of the analysis, as the best possible approximation may be finite with multiple possible values, infinite, or even top, for example, in the case of loop variables or non-determinisms.

When only considering the \num{\TotalEvalScriptsDatasets}~R~scripts for which we could retrieve all read datasets, we get a different distribution of inferred data frame shape constraints, as shown in \Cref{fig:inferred_shapes_datasets}. Relatively speaking, we inferred many more concrete data frame shapes that are not top, and specifically inferred many more finite and exact values compared to the run without all read datasets available.

Across all \num{\TotalEvalSuccess}~analyzed R~scripts, the reduction~\rhoDF resulted in a refinement of the data frame shape in~\qty{\TotalEvalRefinePercent}{\percent}~of all cases where it was applied, underlining the effectiveness of the reduction. In contrast, widening was applies only \num{\TotalEvalWidenings}~times in total. Analyzing the operation arguments that are not data frames, we find that vector arguments were top in~\qty{\TotalEvalVectorArgTopPercent}{\percent}, string arguments in~\qty{\TotalEvalStringArgTopPercent}{\percent}, and number arguments in~\qty{\TotalEvalNumberArgTopPercent}{\percent}~of all cases. This motivates the development of new abstract domains for vectors and numbers in~R in the future.

\begin{finding}
Performing a large-scale evaluation on \num{\TotalEvalScripts}~real-world R~scripts, we inferred \emph{concrete} data frame shapes (that are not top) for~\qty{\TotalEvalDfShapeNotTopPercent}{\percent} and \emph{exact} shapes (with equal lower and upper bounds) for \qty{\TotalEvalDfShapeExactPercent}{\percent}~of the operations. When only considering scripts with all read datasets available, this improves to~\qty{\TotalEvalDatasetsNotTopPercent}{\percent} and \qty{\TotalEvalDatasetsExactPercent}{\percent}, respectively. We see this as a good indication to significantly support researchers in identifying data frame shape issues.
\end{finding}

\paragraph{Invalid Accesses}\label{par:invalid_accesses}
Additionally, we ran our data frame access validation on all \num{\TotalEvalScripts}~real-world R~scripts to identify potential invalid data frame accesses to non-existing columns and rows. Our analysis reported \num{\LinterEvalIssues}~data frame access issues across \num{\LinterEvalIssueScripts}~scripts~(\qty{\LinterEvalIssueScriptsPercent}{\percent}~of all scripts). Of these, \num{\LinterEvalColNamesIssues}~issues were column name access issues~(\qty{\LinterEvalColNamesIssuesPercent}{\percent}), \num{\LinterEvalColIndexIssues}~issues were column index access issues~(\qty{\LinterEvalColIndexIssuesPercent}{\percent}), and there were no row index access issues.

By manually investigating the invalid data frame accesses, we found that the reported issues in \num{\LinterEvalFalsePositiveScripts}~scripts were false positives caused by edge cases that are currently not supported by \flowR or by our implementation, such as placeholders for pipe operators and partial column name matching in accesses.
However, in \num{\LinterEvalActualIssueScripts}~scripts with reported issues, we identified data frame accesses that we assume to be actual data frame access violations. We attribute the invalid data frame access in \num{\LinterEvalActualWrongDatasetScripts}~scripts to a missing column in a read dataset and in \num{\LinterEvalActualWrongCapitalizationScripts}~scripts to the incorrect capitalization of an accessed column name. We assume that the small number of identified data frame access issues is due to the fact that these R~scripts are mostly publication artifacts of peer-reviewed venues and invalid data frame accesses typically result in runtime errors.

\begin{finding}
    Although most of the analyzed R~scripts are from peer-reviewed venues, and invalid data frame accesses typically result in runtime errors, we identified \num{\LinterEvalActualIssueScripts}~real-world R~scripts containing potential actual invalid data frame accesses.
\end{finding}

\paragraph{Data Frame Operations}\label{par:found_operations}
To confirm our findings on \hyperref[par:common_operations]{\textit{Common Operations}} from \cref{sec:approach}, we investigate the most common function calls for which we inferred a data frame shape for an argument or the return value over all analyzed R~scripts. We differentiate between
\begin{andlist}
    \item data frame \emph{sources}, where we did not infer a data frame shape for any function argument but for the return value
    \item data frame \emph{transformations}, where we inferred a data frame shape for at least one argument and for the return value
    \item data frame \emph{sinks}, where we inferred a data frame shape for at least one argument but not for the return value
\end{andlist}.

\Cref{fig:dataframe_functions} shows the \num{10}~most common data frame operations for each of the three categories and the most common operations resulting in an over-approximation. The most common data frame sources are \code{data.frame}, \code{read.csv}, and \code{as.data.frame} which correlates with our findings on \hyperref[par:common_operations]{\textit{Common Operations}} in \cref{sec:approach}. However, the data frame sources also contain data transformation functions like \code{filter}, \code{mutate}, and \code{select}, because we know that they always return data frames, even if we did not detect any argument as data frame. The most common data frame transformation is the pipe operator~\code{\%>\%}, as it just passes along a data frame as implicit first argument to the next function call, followed by the index-based access operation~\code{[} which is used to subset data frames. They are followed by classical data transformation functions like \code{mutate}, \code{filter}, \code{subset}, and \code{select}. The most common data frame sinks are the name-based column access~\code{\$}, the assignment operator~\code{<-}, and name-based column assignment~\code{\$<-}. Besides operations assigning, retrieving, or comparing data frames and their properties, the most common data frame sinks are \code{lm} and \code{ggplot}, motivating future work on checking if data frame shape assumptions of these functions are fulfilled.

To identify possible causes of over-approximations of the data frame shapes, we are also interested in the most common operations where we over-approximate the result with infinite upper bounds~(\topN or~\(+\infty\)). \Cref{fig:overapproximated_functions} shows the \num{10}~most common data frame operations where we inferred a data frame shape with infinite upper bounds for the result. The functions \code{data.frame}~and \code{as.data.frame} are the most common data frame operations whose result was over-approximated with infinite upper bounds. As this is mostly due to the fact that we currently rely on constant propagation for all non-data frame arguments, this highlights the opportunity to improve the data frame shape inference by introducing new static analyses for vectors, lists, and matrices in the future.

\begin{finding}
    The most common data frame source functions are \code{data.frame} and \code{read.csv}, common transformation functions are \code{mutate}, \code{filter}, and \code{subset}, and common sink functions are \code{lm} and \code{ggplot}. The most common functions resulting in an over-approximation are \code{data.frame} and \code{as.data.frame}, motivating future work on abstract domains for vectors, list, and matrices in~R.
\end{finding}

\begin{figure}
    \centering
    \includegraphics[width=\linewidth]{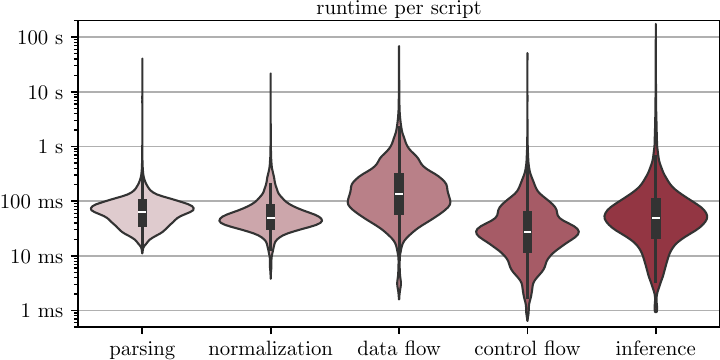}
    \caption{The runtime of the data frame shape inference compared to the parsing, normalization, data flow graph, and control flow graph construction of \flowR in log~scale.}
    \label{fig:benchmark_runtime}
\end{figure}

\paragraph{Runtime}\label{par:runtime}
On a Debian-based virtual server with \num{8}~vCPUs and \qty{8}{\giga\byte}~of RAM, the average runtime of the data frame shape inference over all \num{\TotalEvalSuccess}~analyzed R~scripts was~\(\TotalEvalAvgInferenceTime \pm \TotalEvalSdInferenceTime\) with high outliers and a median runtime of~\TotalEvalMedianInferenceTime. The average runtime of the total analysis including the preceding parsing, abstract syntax tree normalization, data flow graph construction, and control flow graph generation of \flowR was~\(\TotalEvalAvgTotalTime \pm \TotalEvalSdTotalTime\) with a median of~\TotalEvalMedianTotalTime, as shown in \Cref{fig:benchmark_runtime}. We consider an average total runtime of~\TotalEvalAvgTotalTime to be sufficiently fast, as it is within the time range for a user to keep their current flow of thought in an interactive scenario~\cite{ref:books/crc/tucker97/Nielsen97}.

\begin{finding}
    With an average total runtime of~\TotalEvalAvgTotalTime, we consider our analysis to be sufficiently fast for an interactive exploration of inferred data frame shapes.
\end{finding}

\paragraph{\textbf{Summary of \ref{rq2}}}\label{par:rq2_summary}
We empirically verified the soundness of our approach on \num{\CorrEvalScripts}~executable real-world R~scripts by never under-approximating the actual data frame shape. Additionally, we demonstrated the effectiveness of our analysis on a dataset of \num{\TotalEvalScripts}~real-world R~scripts by inferring \emph{concrete} shape constraints that are not top for \qty{\TotalEvalDfShapeNotTopPercent}{\percent}~and \emph{exact} shapes with equal lower and upper bound for~\qty{\TotalEvalDfShapeExactPercent}{\percent}~of the data frame operations, improving to \qty{\TotalEvalDatasetsNotTopPercent}{\percent} and \qty{\TotalEvalDatasetsExactPercent}{\percent}, respectively, when all read datasets are available to our analysis. Using our analysis, we identified \num{\LinterEvalActualIssueScripts}~real-world R~scripts containing potential invalid data frame accesses. We see this as a good indicator to significantly support researchers in working with data frames in data analysis.

\subsection{Threats to Validity}\label{sub:threads}

Threats to the internal validity of our approach are mainly the correctness of our formalization, the implementation, and the data flow analysis of \flowR. Additionally, as we are not aware of any general formal definition of the semantics of~R, our approach is based on our understanding of~R and its semantics. Furthermore, the results of the effectiveness evaluation are based on the assumption that our analysis is sound. We address these threats by our large test suite of \num{\TestsTotal}~systematic unit tests and by verifying the correctness of our analysis based on ground truth from real-world R~scripts.

External threats to the validity of our results include the used limited dataset of real-world R~scripts from publication artifacts that does not include other sources, such as code repositories, notebooks, and R~package code. Nevertheless, we consider our evaluation on \num{\TotalEvalScripts}~real-world R~scripts to be sufficient to demonstrate the effectiveness of our approach on real-world code.

\section{Conclusion}\label{sec:conclusion}

Within this work we find empirical evidence that the shape of data frames can be soundly inferred in real-world R~programs using abstract interpretation, and that this inference can be used to detect potential issues with data frame manipulations. Moreover, these inferred constraints can be treated as static contracts in data science code, even in a highly dynamic language such as~R.

We see three promising directions for future research: First, the inferred data frame shape constraints can be used to derive preconditions on external inputs like read datasets. This enables us to automatically generate contracts for data frames from external inputs and make them explicit, for example, to support reusing a script with a different dataset. Second, we can use the inference to derive optimizations for data frame operations, for example, by rewriting cleaning operations and transformations to more efficient implementations. Third, we can include the content of data frames in the analysis to derive more precise constraints, for example, by tracking the types of values in columns, the relationships between columns, or the statistical properties of the data~\cite{ref:conf/kbse/TurcotteM25}.

\section*{Data Availability}

We provide a publicly available reproduction package for our analysis on Zenodo to verify and reproduce our results~\cite{ref:reproduction-package}.

\printbibliography

\end{document}